\documentclass[%
 reprint,
superscriptaddress,
 amsmath,amssymb,
 aps,
 prl,
]{revtex4-1}

\usepackage{graphicx}
\usepackage{dcolumn}
\usepackage{bm}
\usepackage{xcolor}
\usepackage{hyperref}
\usepackage{kpfonts}
\usepackage{wasysym}
\usepackage{textcomp}

\newcommand*{\citen}[1]{%
  \begingroup
    \romannumeral-`\x 
    \setcitestyle{numbers}%
    \cite{#1}%
  \endgroup   
}

\begin{document}


\title{Adaptable DNA interactions regulate surface triggered self assembly.}

\author{Roberta Lanfranco}
 \affiliation{
 Cavendish Laboratory, University of Cambridge, JJ Thomson Avenue, Cambridge, CB3 0HE, United Kingdom
 }
 
\author{Pritam Kumar Jana}%
\affiliation{
Interdisciplinary Center for Nonlinear Phenomena and Complex Systems, Universit\'{e} libre de Bruxelles (ULB) Campus Plaine, CP 231, Blvd. du Triomphe, B-1050 Brussels, Belgium
}

\author{Gilles Bruylants}
\affiliation{
Engineering of Molecular NanoSystems, Universit\'{e} libre de Bruxelles (ULB), 50 av. F.D. Roosevelt, 1050 Brussels, Belgium
}

\author{Pietro Cicuta}
 \affiliation{
 Cavendish Laboratory, University of Cambridge, JJ Thomson Avenue, Cambridge, CB3 0HE, United Kingdom
 }
 
\author{Bortolo Matteo Mognetti}
\affiliation{
Interdisciplinary Center for Nonlinear Phenomena and Complex Systems, Universit\'{e} libre de Bruxelles (ULB) Campus Plaine, CP 231, Blvd. du Triomphe, B-1050 Brussels, Belgium
}

\author{Lorenzo Di Michele}
  \email{l.di-michele@imperial.ac.uk}
  \affiliation{
  Department of Chemistry, Imperial College London, Molecular Sciences Research Hub, 80 Wood Lane, London W12 0BZ, United Kingdom}
 \affiliation{
 Cavendish Laboratory, University of Cambridge, JJ Thomson Avenue, Cambridge, CB3 0HE, United Kingdom
 }

\date{\today}

\begin{abstract}
DNA-mediated multivalent interactions between colloidal particles have been extensively applied for their ability to program bulk phase behaviour and dynamic processes. Exploiting the competition between different types of DNA-DNA bonds, here we experimentally demonstrate the selective triggering of colloidal self-assembly in the presence of a { functionalised surface, which induces changes in particle-particle interactions}. Besides its relevance to the manufacturing of layered materials with controlled thickness, the intrinsic signal-amplification features of the proposed interaction scheme make it valuable for biosensing applications.

\end{abstract}

\pacs{Valid PACS appear here}
\maketitle

\rmfamily 
Inspired by biology, self-assembly has been extensively investigated for its relevance to the manufacturing of advanced materials. Much of the effort to date has focused on engineering the interactions between nanoscale and colloidal building blocks, with the goal of controlling their bulk phase behaviour and ultimately tailoring the morphological, mechanical, and dynamic features of the resulting materials.~\cite{Seeman:2017aa,Ozin:2009aa,rogers2016using,halverson2013DNA,WangNature2012,Xiong:2020aa}\\
Very few of the biological examples of self-assembly, however, can rightfully be regarded as taking place ``in the bulk''. Indeed, biological macromolecules often operate  in heterogeneous environments or close to functional interfaces, which can trigger self-assembly and regulate the properties and size of the aggregates. For example, Microtubule Organisation Centres such as centrosomes and basal bodies are known to control the self-assembly of microtubules and shape them into morphologically and functionally distinct architectures.\cite{conduit2015centrosome,Pearson:2009aa} While centrosomes sculpt the spindle apparatus, crucial for cell division, basal bodies regulate the emergence of eukaryotic cilia and flagella.~\cite{conduit2015centrosome,Pearson:2009aa} Many more instances of interface-mediated self-assembly can be identified in biology, including the ubiquitous complexation of cell-membrane receptors \cite{wu2013higher} and the self-assembly of viral capsids templated by their nucleic acid cargo.~\cite{Perlmutter:2015aa}\\
Inspired by the recent numerical study of Jana and Mognetti,\cite{Jana2018} here we propose an experimental colloidal system in which self-assembly of finite-size aggregates only occurs in the presence of a functional interface, { while in its absence the colloidal particles exist in a stable gas phase.}  The sought outcome is obtained thanks to the open-ended programmability of DNA-mediated multivalent interactions,\cite{seeman-mirkin-review} and the arsenal of design strategies that ourselves and others have developed to engineer their equilibrium and kinetic features.\cite{stefano-nature,rogers-manoharan,parolini2014thermal,ParoliniACSNano2016,HalversonJCP2016,zhang2017sequential,review2019}
\begin{figure*}[ht!]
\centering
  \includegraphics[width=18cm]{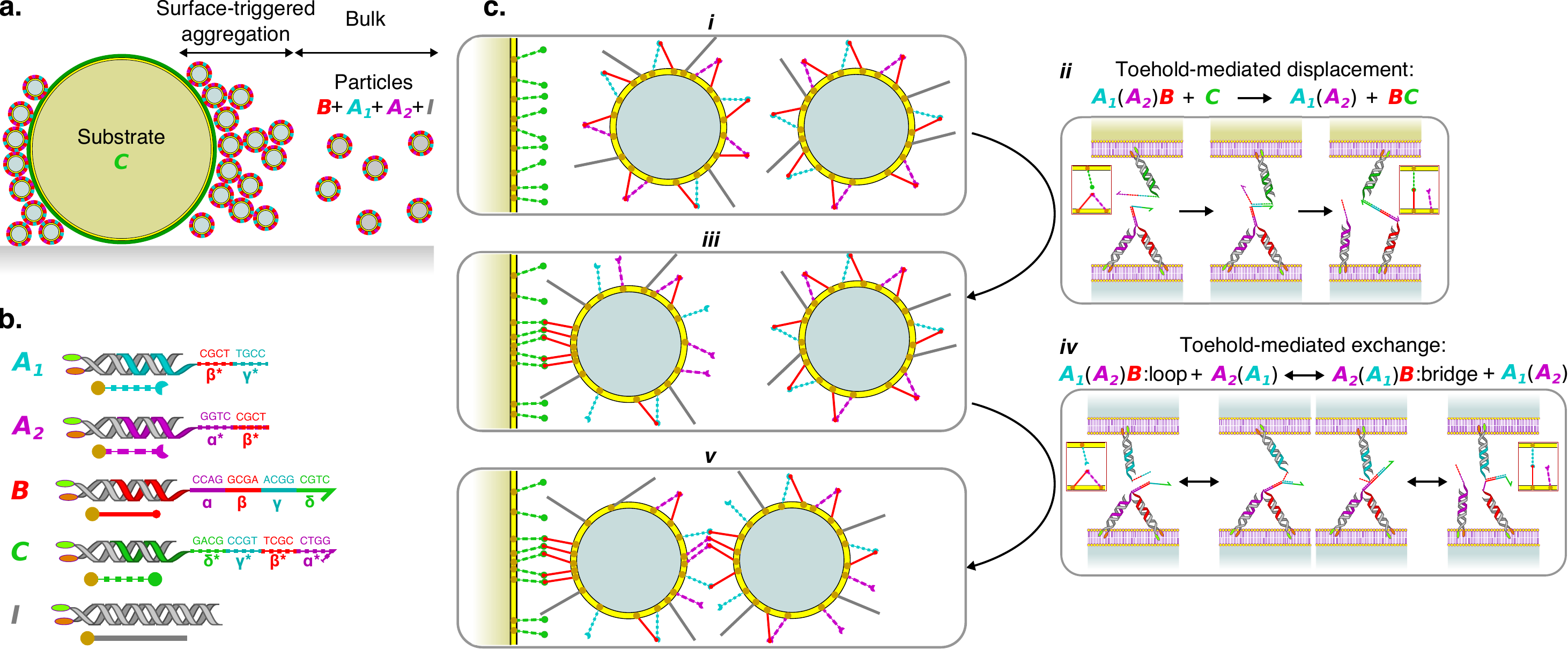}
  \caption{\textbf{Competing DNA bonds and toehold-mediated kinetic control enable programming of surface-triggered assembly.} \textbf{a.} Substrate silica beads ($\diameter \sim 10$\,$\muup$m) and particles ($\diameter  \sim 1$\,$\muup$m) are coated by a SLB. The former are then functionalised with linkers $C$, the latter with linkers $A_1$, $A_2$, $B$ and inert constructs $I$. \textbf{b.} Structure, sticky-end sequence, and domain definition for all the linkers involved in the study. Inert constructs are also shown. \textbf{c.}~Pathway for substrate triggered particle aggregation. \emph{(i)} Particles are thermodynamically stable as a colloidal gas. Nearly all $A_1$, $A_2$ and $B$ linkers are engaged in stable loops $A_1B$ or $A_2B$ loops. \emph{(ii)} Upon interaction of a particle with the substrate, toehold-mediated strand displacement catalyses the breakup of $A_1B$ and $A_2B$ loops and the formation of more stable $BC$ bridges, freeing up $A_1$ and $A_2$ linkers. \emph{(iii)} The particle adheres to the substrate following the formation of several $BC$ bridges and the release of as many $A_1$ or $A_2$ linkers. Partitioning of $B$ and $C$ within the particle-substrate adhesion area and re-distribution of free $A_1$ and $A_2$ is enabled by linker motility. \emph{(iv)} The excess of free $A_1$ and $A_2$ on the particle adhering to the substrate enhances the likelihood of bridge formation with other particles in the bulk, catalysed by toehold-mediated-exchange. \emph{(v)} the formation of particle-particle bridges leads to self-assembly.}
  \label{Figure1}
\end{figure*}

To demonstrate experimentlly this new mechanism, we chose silica colloidal particles with diameter of $\sim$1\,$\muup$m as building blocks. The microspheres are coated with a supported lipid bilayer (SLB), to which DNA constructs (linkers) mediating particle-particle and particle-surface interactions are anchored, as demonstrated in Fig.~\ref{Figure1}\textbf{a}.~\cite{review2019,Troutier:2007aa,VdMeulenJACS2013} DNA linkers feature a 36 base-pair (bp) rigid double-stranded (ds)DNA: the "spacer". One end of the spacer is decorated by two hydrophobic cholesterol/cholesteryl moieties ("anchors"), which cause the linkers to irreversibly insert in the bilayers. Fluorescence-Recovery After Photobleaching (FRAP) measurements, shown in Fig. S1\textbf{a}, demonstrate the fluidity of the SLBs. FRET also confirms that the linkers themselves are capable of undergoing lateral diffusion -- a key characteristic for the designed response of the system (Fig. S1\textbf{b}-\textbf{c}).~\cite{review2019} At the other end of the spacer, linkers feature a single-stranded (ss)DNA "sticky end", the base sequence of which dictates the linker-linker interactions.\\
Much larger silica spheres, with diameter of $\sim10$\,$\muup$m play the role of the trigger surface ("substrate"). Substrate spheres are also coated with a fluid SLB and DNA linkers (Fig.~\ref{Figure1}). Complete information of sample preparation methods and materials are provided in ESI, section S1.\\
As shown in Fig.~\ref{Figure1}\textbf{b}, four types of linkers are present in our system, labelled $A_1$, $A_2$, $B$ and $C$, hosting different sticky ends. Particles feature linkers $A_1$, $A_2$ and $B$, while substrates are only functionalised with $C$ (Fig.~\ref{Figure1}\textbf{a}). In addition to linkers, particles are also decorated with inert constructs $I$, which feature a longer dsDNA spacer (68 bp) and no sticky ends, and are used to regulate steric repulsion~\cite{ParoliniACSNano2016}(Fig.~\ref{Figure1}\textbf{b}).\\
Sticky ends are composed of multiple four-nucleotide (nt) domains, labelled $\alphaup$, $\betaup$, $\gammaup$, $\deltaup$, with their complementary counterparts marked by an asterisk $*$. The sticky ends of linkers $B$ have domain sequence $\alphaup \betaup \gammaup \deltaup$, while those of $A_1$ and $A_2$ feature $ \betaup^*\gammaup^*$ and $\alphaup^*\betaup^*$, respectively. Consequently, $A_1B$ and $A_2B$ bonds can form both between linkers anchored to the same particle (loops) and across different particles (bridges). When an $A_1B$ ($A_2B$) bond is formed, domain $\alphaup$ ($\gammaup$) on $B$ remains accessible and acts as a toehold for sticky end $A_2$ ($A_1$).\cite{zhang2009control} Consequently $A_1B$ and $A_2B$ bonds can readily swap thanks to toehold-mediated exchange, as depicted in Fig.~\ref{Figure1}\textbf{c}, { and previously demonstrated for functionalised liposomes~\cite{ParoliniACSNano2016} and DNA hydrogels.~\cite{Bomboi2019}} The sequences of the sticky ends, along with all DNA oligonucleotides employed in this work are shown in the ESI, Table S1.\\
While the formation of inter-particle bridges induce particle-particle aggregation, a prominence of intra-particle loops tends to stabilise a colloidal gas phase.\cite{BachmannSoftMatter2016}
Free-energy minimisation and {combinatorial entropy demand the coexistence of  inter- and intra-particle bonds.\cite{BachmannSoftMatter2016,DiMicheleJCP2016,HalversonJCP2016,Jana2018,Sciortino2020}} The equilibrium between the two bond-type populations can be systematically tuned by changing the total number of linkers and inert constructs {\em per} particle. Owing to combinatorial considerations, related to the number of ways in which a given type of bond (\emph{e.g.} intra or inter-particle) can be formed, bridges become more favourable when the number of linkers ($A_1$, $A_2$, and $B$) is higher.~\cite{review2019} Instead, inert constructs suppress bridge formation by increasing particle-particle steric repulsion.

\begin{figure}[ht!]
\centering
  \includegraphics[width=9cm]{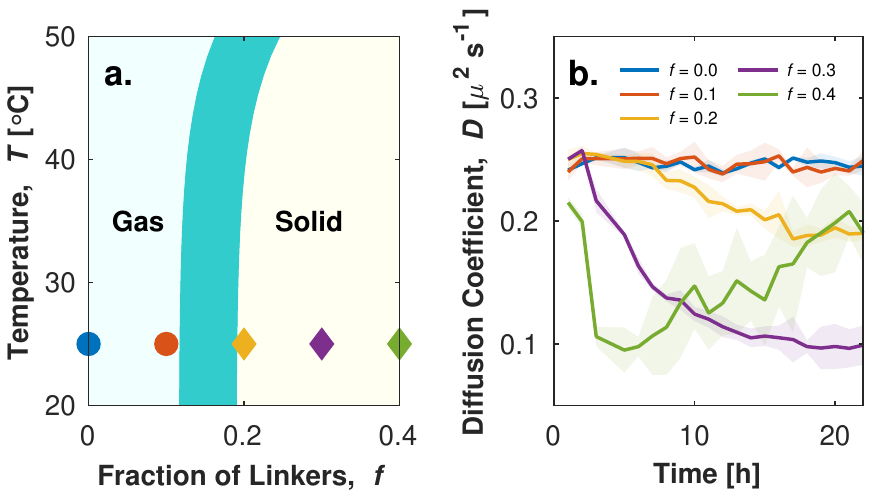}
  \caption{\textbf{Tuning the proportion of linkers enables the stabilisation of a colloidal gas phase}. \textbf{a}. Theoretical phase diagram showing the appearance of a stable colloidal gas phase at sufficiently low fraction of linkers $f$. Full details on the derivation of the phase diagram are provided in ESI, section S2. The finite width of the Gas-Solid phase boundary accounts for the changes in its location following different assumptions in the concentration of the particles and their coordination in the aggregates. Symbols represent conditions tested experimentally and found to display a stable gas phase ($\circ$) or emergence of aggregates ($\diamond$), as determined from diffusivity data and visual inspection (see ESI Fig. S2). \textbf{b}.~Time evolution of the effective diffusion coefficient $D$ of initially isolated particles, as determined with DDM.~\cite{Cerbino:2008aa,Cerbino:2017aa} A drop in $D$ indicates particle aggregation. 
  Shadowed regions mark the errorbars of the curves calculated as discussed in the ESI, Section S1.2.1.}
  \label{Figure2}
\end{figure}

{ At first, we seek to identify experimental conditions under which a colloidal gas phase is stable in the bulk. We will then prove that, under these conditions,} the presence of the substrates can trigger the self-assembly of finite-size particle aggregates on their surface.\\
Thanks to a comprehensive theoretical framework recently summarised in Ref. \citen{review2019}, we can predict particle-particle interaction potentials and their bulk phase behaviour, as detailed in ESI section S2, and thus guide experiment design. We consider the experimentally-relevant scenario in which particles are functionalised with $1.6\times 10^5$ constructs, including sticky linkers and inert constructs. The concentration ratios of the three linker species are selected such that $[A_1]=[A_2]=0.5\times[B]$. We define the relative concentration of linkers over the total concentration of particle-tethered constructs as $f=[L]/([L]+[I])$, with $[L]=[A_1]+[A_2]+[B]$.\\
The diagram in Fig.~\ref{Figure2}\textbf{a} maps the bulk phase behaviour of the particles as a function of $f$ and the temperature $T$.
A stable colloidal gas phase is predicted at room temperature ($T=25^\circ$C) for $f<0.15 \pm 0.04$, while aggregates emerge at greater fractions of linkers. The errorbar in the phase boundary derives from different assumptions in particle concentration and their coordination within the aggregates (see ESI section S2).
At sufficiently low $f$ the particle-particle steric repulsion induced by the inert constructs suppresses the formation of inter-particle bridges. Since $[B]=[A_1]+[A_2]$, in this regime and at sufficiently low temperature, all the available linkers are engaged in loops.\\

To experimentally confirm the predicted bulk phase behaviour of the particles we prepared samples with $f=0, 0.1, 0.2,0.3$ and  0.4, and studied their aggregation (or lack thereof) \emph{via} bright field light microscopy. Following a strategy developed in Ref.~\citen{ParoliniACSNano2016}, particles are initially forced in a state in which only loops can be formed by performing a rapid temperature quench (see ESI section S1.1.3). Since no bridges are present, samples exhibit a gas phase, which may be either stable or metastable depending on $f$ and $T$. The samples are then observed over time at room temperature ($T\simeq25^\circ$C). In conditions where bridge formation is thermodynamically feasible, namely at sufficiently high $f$, aggregation is expected to occur. Note that, as discussed above, loop breakup and bridge formation is kinetically aided by the built-in toehold-mediated-exchange capabilities of the system.\cite{zhang2009control,ParoliniACSNano2016} To quantitatively assess the occurrence of particle-particle aggregation we carry out Differential Dynamic Microscopy (DDM) measurements, which allow us to determine the time-evolution of the (apparent) diffusion coefficient $D$ of individual particles and possible aggregates~\cite{Cerbino:2008aa,Cerbino:2017aa} (see ESI section S1.2.1). Data are summarised, for different values of $f$, in Fig.~\ref{Figure2}\textbf{b} (see also ESI Fig. S2\textbf{a}).
Consistently with theoretical predictions in Fig.~\ref{Figure2}\textbf{a}, $D$ displays a decreasing trend with time for $f\ge0.2$, following the emergence of aggregates. While for $f=0.2$ $D$ exhibits a slight decrease, indicative of small and sparse aggregates (ESI Fig. S1\textbf{b} iii), the trend becomes more prominent for $f=0.3$ where we observe large scale aggregation (ESI Fig. S2\textbf{b} iv). Samples with $f=0.4$ display a sharp drop in diffusion coefficient from the beginning of the observation, followed by a slight increase at later times. The latter trend follows from the formation of very large aggregates (ESI Fig. S2\textbf{b} v); these are largely immobile and contribute little to the DDM signal, which is in turn dominated by the few smaller aggregates that remain diffusive. { The complex fluctuation dynamics of branched colloidal aggregates may also play a role in the upturn of $D$ (Fig. S2\textbf{a}).~\cite{Cho:2020aa}} For $f=0.1$,  $D$ remains constant over a 22-hour observation window, indicating lack of particle aggregation (ESI Fig. S2\textbf{b} ii) and the stability of the colloidal gas phase.\\ 

\begin{figure}[t!]
\centering
  \includegraphics[width=9cm]{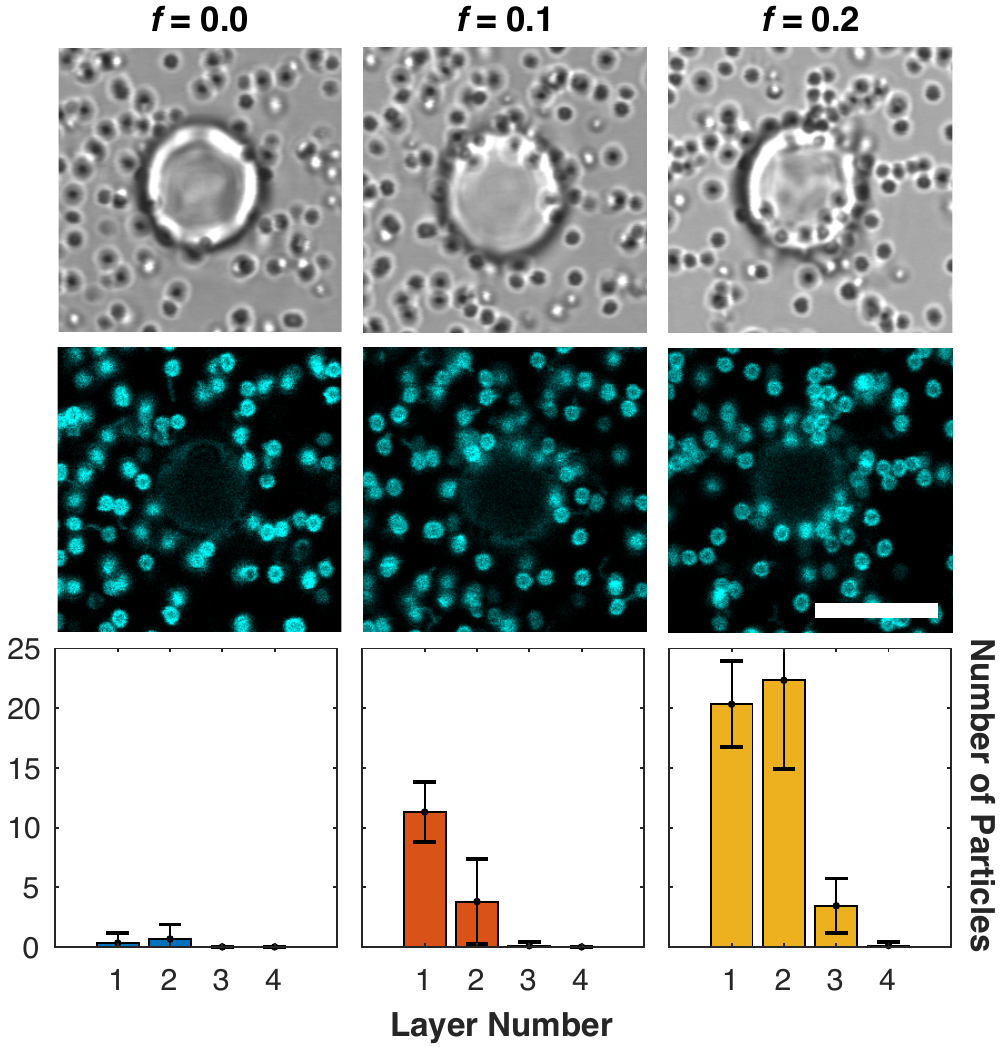}
  \caption{\textbf{Fraction of linkers and inert constructs regulates the growth of surface-triggered aggregates}. Bright-field images (Top) and confocal cross-sections (Middle) of a typical substrate sphere and surrounding particles for relevant values of the fraction of linkers $f$. Particle adhesion and substrate driven aggregation are observed for $f=0.1$, although no aggregate formation is detected in bulk for this condition (Fig.~\ref{Figure2}\textbf{b}). Bottom: histograms quantifying the average numbers of particles per substrate bead for each "adhesion layer", with the particles directly adhering to the substrate classified as belonging to layer 1. See ESI section S.1.2.2 for the definition of the layers and the particle-classification procedure. Scale bars: $10$\,$\muup$m.}
  \label{Figure3}
\end{figure}

Having identified the sought conditions, under which particles do not aggregate in bulk, we introduce substrate spheres functionalised with $C$ linkers. The sticky ends of $C$ feature domain sequence $\deltaup^*\gammaup^*\betaup^*\alphaup^*$, and are thus fully complementary to $B$ sticky ends. As a result $BC$ bridges between a particle and a substrate are expected to be substantially more stable than either $A_1B$ or $A_2B$ bridges. We can thus expect $BC$ bridge formation and adhesion of particles to substrates even for samples with stable bulk gas phase ($f=0.1$), as sketched in Fig.\ref{Figure1}\textbf{c} i-iii. Note that the breakup of $A_1B$ or $A_2B$ loops initially present on the particle in favour of $BC$ bridges is made kinetically accessible by toehold-mediated strand displacement \cite{zhang2009control} (Fig.\ref{Figure1}\textbf{c} ii). Once formed, $BC$ bridges are effectively irreversible under experimental conditions, given the total length of the complementary domains and that they do not offer any toehold to $A_1$ or $A_2$.\\
For every $BC$ bridge that forms, either a $A_1$ or a $A_2$ linker initially engaged in a loop becomes free. The lateral mobility of all linkers implies that loop breakup is not limited to the small area of contact between the substrate and the particles. Instead, a large number of bridges can be formed, following the recruitment of $B$ and $C$ linkers in the contact area (Fig.\ref{Figure1}\textbf{c} iii).~\cite{parolini2014thermal,LanfrancoLangmuir2019} At the same time, the newly unbound $A_1$ and $A_2$ linkers are free to uniformly spread on the surface of the particle, now adhering to the substrate (Fig.\ref{Figure1}\textbf{c} iii).
As a result, particles adhering to the substrates feature a significantly higher number of unbound $A_1$ and $A_2$ linkers compared to those in bulk. Straightforward combinatorial reasoning prescribes that such an excess of free linkers should facilitate the formation of bridges between adhering particles and those left in the bulk, triggering the deposition of a second particle, as depicted in Fig.\ref{Figure1}\textbf{c} iii-v.\\
Particles in this "second layer" will also display an excess of free $A_1$ and $A_2$, following from the formation of $A_1B$ and $A_2B$ bridges with the particle initially adhering to the substrate, which could trigger the adhesion of a third particle. This amplification process can lead to the surface-triggered growth of aggregates, as per our objective.\\
{ In other words, the presence of the substrate leads to a dynamic adaptation of particle-particle interactions that propagate also to particles which are not in direct contact with the functional interface.}
Numerical calculations performed by Jana and Mognetti on a closely related system \cite{Jana2018} demonstrate how the number of linkers freed up following bridge formation, { and thus the free energy gain of particle-particle adhesion,} decreases with the number of layers, to a point { where} it is no longer sufficient to stabilise further adhesion of the bulk particles. Hence, the growth of aggregates is expected to be self-limiting, { and the size of the aggregates dependent of system parameters such as the relative strength of $BC$ binding, the surface density of linker $C$ and the bulk concentration of particles~\cite{Jana2018}.}\\

Experimental results on surface-triggered aggregate growth are summarised in Fig.~\ref{Figure3}. Confocal and bright-field micrographs demonstrate the presence and morphology of the aggregates, or their absence. Three-dimensional confocal stacks were further analysed to quantify, for each substrate sphere, the average number of particles in each "layer" of adhesion, as detailed in ESI and summarised by the histograms in Fig.~\ref{Figure3} (bottom).\\
As expected, for $f=0$ no aggregation on the substrate particles is observed, besides a minimal degree of adhesion due to non-specific interactions. For $f=0.1$  we observe a significant degree of particle adhesion, and importantly the presence of particles in the second layer of adhesion, confirming the sought effect of surface-triggered assembly of finite-size structures. Samples with $f=0.2$, which display only marginal bulk aggregation (ESI Fig. S3\textbf{b} iii), form instead large particle assemblies on the substrates with a third and even a fourth adhesion layer being observed. Since these large aggregates ultimately originate from a small number of $C$ linker molecules, which produce a larger number of  bridges between particles in subsequent layers, our system offers a mechanism for amplifying and visualising molecular signals.\\
{ Similar to bulk particle aggregation (Fig.~\ref{Figure2}\textbf{b} and Fig. S3, ESI), surface-triggered clustering develops over the course of a few hours, likely limited by the rate of bond-swapping~\cite{ParoliniACSNano2016} and the diffusivity of the relatively large colloidal particles.}\\
As a control, DDM was also performed on samples containing substrate beads. Since large beads are effectively immobile over the timescales of the DDM observations, and so are the particles adhering to them, we expect this technique to only be sensitive to diffusive particles and aggregates in the bulk. Data in Fig.~S3 show trends very similar to those measured for particle-only samples (Fig.~\ref{Figure2}\textbf{b}), demonstrating that substrate spheres do not affect the bulk behaviour of the system in unwanted ways (\emph{e.g.} by releasing linkers in solution), but only act locally regulating the growth of aggregates.\\

In summary, with this communication we demonstrate the rational design and experimental implementation of a colloidal system in which self-assembly is controlled by a functional interface. The latter triggers the formation of aggregates which are otherwise unstable in the bulk phase, while also limiting their final size. The degree of control and of our proof-of-concept system derives from the thermodynamic and kinetic programmability of DNA-DNA base-pairing, and our deep understanding of multivalent DNA-mediated interactions.\cite{seeman-mirkin-review,rogers2016using,review2019}\\
Our proof-of-concept system mimics key features of the spatially regulated assembly observed in biology, and we argue that similar approaches could be implemented in more complex biomimicry, including engineering the collective behaviour of synthetic cellular systems.~\cite{Buddingh:2017aa}\\ 
Furthermore, the assembly of large colloidal aggregates, as triggered by a relatively small number of nucleic acid molecules ($C$ linkers) could be used as an amplification and detection system for nucleic-acid biomarkers, such as disease related microRNA, ciruclating DNA or viral genetic material.~\cite{Vaart:2007aa,Li:2009aa,Mackay:2002aa,OBrien:2018aa} Such a response could be obtained by re-desiging $A_1$, $A_2$ and $B$ sticky ends to respond to a new target sequence different from $C$, then immobilising the (possibly present) target on substrate beads, and checking for the occurrence of substrate-triggered particle aggregation as an amplified readout. { Immobilisation could be achieved by pre-functionalising substrate spheres with cholesterolised linkers partially complementary to the target strands. The timescales of aggregate formation could be accelerated for faster readout, \emph{e.g.} by using smaller, faster diffusing particles, or slightly increasing the length of the toehold domains to improve the rate of bond-swapping.~\cite{zhang2009control}}\\

\begin{acknowledgments}
\textbf{Acknowledgments}\\
RL, BMM, PC, LDM, and GB acknowledge support
from the Wiener-Anspach Foundation. PKJ and BMM acknowledge support from Fonds de la Recherche Scientifique de Belgique (F.R.S.-FNRS) under grant n$^\circ$ MIS F.4534.17 and by an ARC (ULB) grant of the {\em F\'ed\'eration Wallonie-Bruxelles}. LDM acknowledges support from a Royal Society University Research Fellowship (UF160152) and from the European Research Council (ERC) under the Horizon 2020 Research and Innovation Programme (ERC-STG No 851667 NANOCELL).
\end{acknowledgments}


\begin{thebibliography}{35}%
\makeatletter
\providecommand \@ifxundefined [1]{%
 \@ifx{#1\undefined}
}%
\providecommand \@ifnum [1]{%
 \ifnum #1\expandafter \@firstoftwo
 \else \expandafter \@secondoftwo
 \fi
}%
\providecommand \@ifx [1]{%
 \ifx #1\expandafter \@firstoftwo
 \else \expandafter \@secondoftwo
 \fi
}%
\providecommand \natexlab [1]{#1}%
\providecommand \enquote  [1]{``#1''}%
\providecommand \bibnamefont  [1]{#1}%
\providecommand \bibfnamefont [1]{#1}%
\providecommand \citenamefont [1]{#1}%
\providecommand \href@noop [0]{\@secondoftwo}%
\providecommand \href [0]{\begingroup \@sanitize@url \@href}%
\providecommand \@href[1]{\@@startlink{#1}\@@href}%
\providecommand \@@href[1]{\endgroup#1\@@endlink}%
\providecommand \@sanitize@url [0]{\catcode `\\12\catcode `\$12\catcode
  `\&12\catcode `\#12\catcode `\^12\catcode `\_12\catcode `\%12\relax}%
\providecommand \@@startlink[1]{}%
\providecommand \@@endlink[0]{}%
\providecommand \url  [0]{\begingroup\@sanitize@url \@url }%
\providecommand \@url [1]{\endgroup\@href {#1}{\urlprefix }}%
\providecommand \urlprefix  [0]{URL }%
\providecommand \Eprint [0]{\href }%
\providecommand \doibase [0]{http://dx.doi.org/}%
\providecommand \selectlanguage [0]{\@gobble}%
\providecommand \bibinfo  [0]{\@secondoftwo}%
\providecommand \bibfield  [0]{\@secondoftwo}%
\providecommand \translation [1]{[#1]}%
\providecommand \BibitemOpen [0]{}%
\providecommand \bibitemStop [0]{}%
\providecommand \bibitemNoStop [0]{.\EOS\space}%
\providecommand \EOS [0]{\spacefactor3000\relax}%
\providecommand \BibitemShut  [1]{\csname bibitem#1\endcsname}%
\let\auto@bib@innerbib\@empty
\bibitem [{\citenamefont {Seeman}\ and\ \citenamefont
  {Sleiman}(2017)}]{Seeman:2017aa}%
  \BibitemOpen
  \bibfield  {author} {\bibinfo {author} {\bibfnamefont {N.~C.}\ \bibnamefont
  {Seeman}}\ and\ \bibinfo {author} {\bibfnamefont {H.~F.}\ \bibnamefont
  {Sleiman}},\ }\href@noop {} {\bibfield  {journal} {\bibinfo  {journal} {Nat.
  Rev. Mater.}\ }\textbf {\bibinfo {volume} {3}},\ \bibinfo {pages} {17068}
  (\bibinfo {year} {2017})}\BibitemShut {NoStop}%
\bibitem [{\citenamefont {Ozin}\ \emph {et~al.}(2009)\citenamefont {Ozin},
  \citenamefont {Hou}, \citenamefont {Lotsch}, \citenamefont {Cademartiri},
  \citenamefont {Puzzo}, \citenamefont {Scotognella}, \citenamefont {Ghadimi},\
  and\ \citenamefont {Thomson}}]{Ozin:2009aa}%
  \BibitemOpen
  \bibfield  {author} {\bibinfo {author} {\bibfnamefont {G.~A.}\ \bibnamefont
  {Ozin}}, \bibinfo {author} {\bibfnamefont {K.}~\bibnamefont {Hou}}, \bibinfo
  {author} {\bibfnamefont {B.~V.}\ \bibnamefont {Lotsch}}, \bibinfo {author}
  {\bibfnamefont {L.}~\bibnamefont {Cademartiri}}, \bibinfo {author}
  {\bibfnamefont {D.~P.}\ \bibnamefont {Puzzo}}, \bibinfo {author}
  {\bibfnamefont {F.}~\bibnamefont {Scotognella}}, \bibinfo {author}
  {\bibfnamefont {A.}~\bibnamefont {Ghadimi}}, \ and\ \bibinfo {author}
  {\bibfnamefont {J.}~\bibnamefont {Thomson}},\ }\href {\doibase
  https://doi.org/10.1016/S1369-7021(09)70156-7} {\bibfield  {journal}
  {\bibinfo  {journal} {Mater. Today}\ }\textbf {\bibinfo {volume} {12}},\
  \bibinfo {pages} {12} (\bibinfo {year} {2009})}\BibitemShut {NoStop}%
\bibitem [{\citenamefont {Rogers}\ \emph {et~al.}(2016)\citenamefont {Rogers},
  \citenamefont {Shih},\ and\ \citenamefont {Manoharan}}]{rogers2016using}%
  \BibitemOpen
  \bibfield  {author} {\bibinfo {author} {\bibfnamefont {W.~B.}\ \bibnamefont
  {Rogers}}, \bibinfo {author} {\bibfnamefont {W.~M.}\ \bibnamefont {Shih}}, \
  and\ \bibinfo {author} {\bibfnamefont {V.~N.}\ \bibnamefont {Manoharan}},\
  }\href {\doibase 10.1038/natrevmats.2016.8} {\bibfield  {journal} {\bibinfo
  {journal} {Nat. Rev. Mater.}\ }\textbf {\bibinfo {volume} {1}},\ \bibinfo
  {pages} {16008} (\bibinfo {year} {2016})}\BibitemShut {NoStop}%
\bibitem [{\citenamefont {Halverson}\ and\ \citenamefont
  {Tkachenko}(2013)}]{halverson2013DNA}%
  \BibitemOpen
  \bibfield  {author} {\bibinfo {author} {\bibfnamefont {J.~D.}\ \bibnamefont
  {Halverson}}\ and\ \bibinfo {author} {\bibfnamefont {A.~V.}\ \bibnamefont
  {Tkachenko}},\ }\href@noop {} {\bibfield  {journal} {\bibinfo  {journal}
  {Phys. Rev. E}\ }\textbf {\bibinfo {volume} {87}},\ \bibinfo {pages} {062310}
  (\bibinfo {year} {2013})}\BibitemShut {NoStop}%
\bibitem [{\citenamefont {Wang}\ \emph {et~al.}(2012)\citenamefont {Wang},
  \citenamefont {Wang}, \citenamefont {R.}, \citenamefont {Manoharan},
  \citenamefont {Feng}, \citenamefont {Hollingsworth}, \citenamefont {Weck},\
  and\ \citenamefont {Pine}}]{WangNature2012}%
  \BibitemOpen
  \bibfield  {author} {\bibinfo {author} {\bibfnamefont {Y.}~\bibnamefont
  {Wang}}, \bibinfo {author} {\bibfnamefont {Y.}~\bibnamefont {Wang}}, \bibinfo
  {author} {\bibfnamefont {D.}~\bibnamefont {R.}}, \bibinfo {author}
  {\bibfnamefont {V.~N.}\ \bibnamefont {Manoharan}}, \bibinfo {author}
  {\bibfnamefont {L.}~\bibnamefont {Feng}}, \bibinfo {author} {\bibfnamefont
  {A.~D.}\ \bibnamefont {Hollingsworth}}, \bibinfo {author} {\bibfnamefont
  {M.}~\bibnamefont {Weck}}, \ and\ \bibinfo {author} {\bibfnamefont {D.~J.}\
  \bibnamefont {Pine}},\ }\href {http://dx.doi.org/10.1038/nature11564}
  {\bibfield  {journal} {\bibinfo  {journal} {Nature}\ }\textbf {\bibinfo
  {volume} {491}},\ \bibinfo {pages} {51} (\bibinfo {year} {2012})}\BibitemShut
  {NoStop}%
\bibitem [{\citenamefont {Xiong}\ \emph {et~al.}(2020)\citenamefont {Xiong},
  \citenamefont {Yang}, \citenamefont {Tian}, \citenamefont {Michelson},
  \citenamefont {Xiang}, \citenamefont {Xin},\ and\ \citenamefont
  {Gang}}]{Xiong:2020aa}%
  \BibitemOpen
  \bibfield  {author} {\bibinfo {author} {\bibfnamefont {Y.}~\bibnamefont
  {Xiong}}, \bibinfo {author} {\bibfnamefont {S.}~\bibnamefont {Yang}},
  \bibinfo {author} {\bibfnamefont {Y.}~\bibnamefont {Tian}}, \bibinfo {author}
  {\bibfnamefont {A.}~\bibnamefont {Michelson}}, \bibinfo {author}
  {\bibfnamefont {S.}~\bibnamefont {Xiang}}, \bibinfo {author} {\bibfnamefont
  {H.}~\bibnamefont {Xin}}, \ and\ \bibinfo {author} {\bibfnamefont
  {O.}~\bibnamefont {Gang}},\ }\bibfield  {booktitle} {\emph {\bibinfo
  {booktitle} {ACS Nano}},\ }\href {\doibase 10.1021/acsnano.0c00607}
  {\bibfield  {journal} {\bibinfo  {journal} {ACS Nano}\ ,\ \bibinfo {pages}
  {10.1021/acsnano.0c00607}} (\bibinfo {year} {2020})}\BibitemShut {NoStop}%
\bibitem [{\citenamefont {Conduit}\ \emph {et~al.}(2015)\citenamefont
  {Conduit}, \citenamefont {Wainman},\ and\ \citenamefont
  {Raff}}]{conduit2015centrosome}%
  \BibitemOpen
  \bibfield  {author} {\bibinfo {author} {\bibfnamefont {P.~T.}\ \bibnamefont
  {Conduit}}, \bibinfo {author} {\bibfnamefont {A.}~\bibnamefont {Wainman}}, \
  and\ \bibinfo {author} {\bibfnamefont {J.~W.}\ \bibnamefont {Raff}},\
  }\href@noop {} {\bibfield  {journal} {\bibinfo  {journal} {Nat. Rev. Mol.
  Cell Biol.}\ }\textbf {\bibinfo {volume} {16}},\ \bibinfo {pages} {611}
  (\bibinfo {year} {2015})}\BibitemShut {NoStop}%
\bibitem [{\citenamefont {Pearson}\ and\ \citenamefont
  {Winey}(2009)}]{Pearson:2009aa}%
  \BibitemOpen
  \bibfield  {author} {\bibinfo {author} {\bibfnamefont {C.~G.}\ \bibnamefont
  {Pearson}}\ and\ \bibinfo {author} {\bibfnamefont {M.}~\bibnamefont
  {Winey}},\ }\bibfield  {booktitle} {\emph {\bibinfo {booktitle} {Traffic}},\
  }\href {\doibase 10.1111/j.1600-0854.2009.00885.x} {\bibfield  {journal}
  {\bibinfo  {journal} {Traffic}\ }\textbf {\bibinfo {volume} {10}},\ \bibinfo
  {pages} {461} (\bibinfo {year} {2009})}\BibitemShut {NoStop}%
\bibitem [{\citenamefont {Wu}(2013)}]{wu2013higher}%
  \BibitemOpen
  \bibfield  {author} {\bibinfo {author} {\bibfnamefont {H.}~\bibnamefont
  {Wu}},\ }\href@noop {} {\bibfield  {journal} {\bibinfo  {journal} {Cell}\
  }\textbf {\bibinfo {volume} {153}},\ \bibinfo {pages} {287} (\bibinfo {year}
  {2013})}\BibitemShut {NoStop}%
\bibitem [{\citenamefont {Perlmutter}\ and\ \citenamefont
  {Hagan}(2015)}]{Perlmutter:2015aa}%
  \BibitemOpen
  \bibfield  {author} {\bibinfo {author} {\bibfnamefont {J.~D.}\ \bibnamefont
  {Perlmutter}}\ and\ \bibinfo {author} {\bibfnamefont {M.~F.}\ \bibnamefont
  {Hagan}},\ }\href {\doibase 10.1146/annurev-physchem-040214-121637}
  {\bibfield  {journal} {\bibinfo  {journal} {Annu. Rev. Phys. Chem.}\ }\textbf
  {\bibinfo {volume} {66}},\ \bibinfo {pages} {217} (\bibinfo {year}
  {2015})}\BibitemShut {NoStop}%
\bibitem [{\citenamefont {Jana}\ and\ \citenamefont
  {Mognetti}(2019)}]{Jana2018}%
  \BibitemOpen
  \bibfield  {author} {\bibinfo {author} {\bibfnamefont {P.~K.}\ \bibnamefont
  {Jana}}\ and\ \bibinfo {author} {\bibfnamefont {B.~M.}\ \bibnamefont
  {Mognetti}},\ }\href@noop {} {\bibfield  {journal} {\bibinfo  {journal}
  {Nanoscale}\ }\textbf {\bibinfo {volume} {11}},\ \bibinfo {pages} {5450}
  (\bibinfo {year} {2019})}\BibitemShut {NoStop}%
\bibitem [{\citenamefont {Jones}\ \emph {et~al.}(2015)\citenamefont {Jones},
  \citenamefont {Seeman},\ and\ \citenamefont {Mirkin}}]{seeman-mirkin-review}%
  \BibitemOpen
  \bibfield  {author} {\bibinfo {author} {\bibfnamefont {M.~R.}\ \bibnamefont
  {Jones}}, \bibinfo {author} {\bibfnamefont {N.~C.}\ \bibnamefont {Seeman}}, \
  and\ \bibinfo {author} {\bibfnamefont {C.~A.}\ \bibnamefont {Mirkin}},\
  }\href@noop {} {\bibfield  {journal} {\bibinfo  {journal} {Science}\ }\textbf
  {\bibinfo {volume} {347}},\ \bibinfo {pages} {1260901} (\bibinfo {year}
  {2015})}\BibitemShut {NoStop}%
\bibitem [{\citenamefont {{ Angioletti-Uberti}}\ \emph
  {et~al.}(2012)\citenamefont {{ Angioletti-Uberti}}, \citenamefont
  {Mognetti},\ and\ \citenamefont {Frenkel}}]{stefano-nature}%
  \BibitemOpen
  \bibfield  {author} {\bibinfo {author} {\bibfnamefont {S.}~\bibnamefont {{
  Angioletti-Uberti}}}, \bibinfo {author} {\bibfnamefont {B.}~\bibnamefont
  {Mognetti}}, \ and\ \bibinfo {author} {\bibfnamefont {D.}~\bibnamefont
  {Frenkel}},\ }\href {\doibase
  http://www.nature.com/nmat/journal/v11/n6/full/nmat3314.html} {\bibfield
  {journal} {\bibinfo  {journal} {Nat. Mater.}\ }\textbf {\bibinfo {volume}
  {11}},\ \bibinfo {pages} {518} (\bibinfo {year} {2012})}\BibitemShut
  {NoStop}%
\bibitem [{\citenamefont {Rogers}\ and\ \citenamefont
  {Manoharan}(2015)}]{rogers-manoharan}%
  \BibitemOpen
  \bibfield  {author} {\bibinfo {author} {\bibfnamefont {W.~B.}\ \bibnamefont
  {Rogers}}\ and\ \bibinfo {author} {\bibfnamefont {V.~N.}\ \bibnamefont
  {Manoharan}},\ }\href@noop {} {\bibfield  {journal} {\bibinfo  {journal}
  {Science}\ }\textbf {\bibinfo {volume} {347}},\ \bibinfo {pages} {639}
  (\bibinfo {year} {2015})}\BibitemShut {NoStop}%
\bibitem [{\citenamefont {Parolini}\ \emph {et~al.}(2015)\citenamefont
  {Parolini}, \citenamefont {Mognetti}, \citenamefont {Kotar}, \citenamefont
  {Eiser}, \citenamefont {Cicuta},\ and\ \citenamefont
  {Di~Michele}}]{parolini2014thermal}%
  \BibitemOpen
  \bibfield  {author} {\bibinfo {author} {\bibfnamefont {L.}~\bibnamefont
  {Parolini}}, \bibinfo {author} {\bibfnamefont {B.~M.}\ \bibnamefont
  {Mognetti}}, \bibinfo {author} {\bibfnamefont {J.}~\bibnamefont {Kotar}},
  \bibinfo {author} {\bibfnamefont {E.}~\bibnamefont {Eiser}}, \bibinfo
  {author} {\bibfnamefont {P.}~\bibnamefont {Cicuta}}, \ and\ \bibinfo {author}
  {\bibfnamefont {L.}~\bibnamefont {Di~Michele}},\ }\href@noop {} {\bibfield
  {journal} {\bibinfo  {journal} {Nat. Comm.}\ }\textbf {\bibinfo {volume}
  {6}},\ \bibinfo {pages} {5948} (\bibinfo {year} {2015})}\BibitemShut
  {NoStop}%
\bibitem [{\citenamefont {Parolini}\ \emph {et~al.}(2016)\citenamefont
  {Parolini}, \citenamefont {Kotar}, \citenamefont {Di~Michele},\ and\
  \citenamefont {Mognetti}}]{ParoliniACSNano2016}%
  \BibitemOpen
  \bibfield  {author} {\bibinfo {author} {\bibfnamefont {L.}~\bibnamefont
  {Parolini}}, \bibinfo {author} {\bibfnamefont {J.}~\bibnamefont {Kotar}},
  \bibinfo {author} {\bibfnamefont {L.}~\bibnamefont {Di~Michele}}, \ and\
  \bibinfo {author} {\bibfnamefont {B.~M.}\ \bibnamefont {Mognetti}},\
  }\href@noop {} {\bibfield  {journal} {\bibinfo  {journal} {ACS Nano}\
  }\textbf {\bibinfo {volume} {10}},\ \bibinfo {pages} {2392} (\bibinfo {year}
  {2016})}\BibitemShut {NoStop}%
\bibitem [{\citenamefont {Halverson}\ and\ \citenamefont
  {Tkachenko}(2016)}]{HalversonJCP2016}%
  \BibitemOpen
  \bibfield  {author} {\bibinfo {author} {\bibfnamefont {J.~D.}\ \bibnamefont
  {Halverson}}\ and\ \bibinfo {author} {\bibfnamefont {A.~V.}\ \bibnamefont
  {Tkachenko}},\ }\href@noop {} {\bibfield  {journal} {\bibinfo  {journal} {J.
  Chem. Phys.}\ }\textbf {\bibinfo {volume} {144}},\ \bibinfo {pages} {094903}
  (\bibinfo {year} {2016})}\BibitemShut {NoStop}%
\bibitem [{\citenamefont {Zhang}\ \emph {et~al.}(2017)\citenamefont {Zhang},
  \citenamefont {McMullen}, \citenamefont {Pontani}, \citenamefont {He},
  \citenamefont {Sha}, \citenamefont {Seeman}, \citenamefont {Brujic},\ and\
  \citenamefont {Chaikin}}]{zhang2017sequential}%
  \BibitemOpen
  \bibfield  {author} {\bibinfo {author} {\bibfnamefont {Y.}~\bibnamefont
  {Zhang}}, \bibinfo {author} {\bibfnamefont {A.}~\bibnamefont {McMullen}},
  \bibinfo {author} {\bibfnamefont {L.-L.}\ \bibnamefont {Pontani}}, \bibinfo
  {author} {\bibfnamefont {X.}~\bibnamefont {He}}, \bibinfo {author}
  {\bibfnamefont {R.}~\bibnamefont {Sha}}, \bibinfo {author} {\bibfnamefont
  {N.~C.}\ \bibnamefont {Seeman}}, \bibinfo {author} {\bibfnamefont
  {J.}~\bibnamefont {Brujic}}, \ and\ \bibinfo {author} {\bibfnamefont {P.~M.}\
  \bibnamefont {Chaikin}},\ }\href@noop {} {\bibfield  {journal} {\bibinfo
  {journal} {Nat. Comm.}\ }\textbf {\bibinfo {volume} {8}},\ \bibinfo {pages}
  {21} (\bibinfo {year} {2017})}\BibitemShut {NoStop}%
\bibitem [{\citenamefont {Mognetti}\ \emph {et~al.}(2019)\citenamefont
  {Mognetti}, \citenamefont {Cicuta},\ and\ \citenamefont
  {Di~Michele}}]{review2019}%
  \BibitemOpen
  \bibfield  {author} {\bibinfo {author} {\bibfnamefont {B.~M.}\ \bibnamefont
  {Mognetti}}, \bibinfo {author} {\bibfnamefont {P.}~\bibnamefont {Cicuta}}, \
  and\ \bibinfo {author} {\bibfnamefont {L.}~\bibnamefont {Di~Michele}},\
  }\bibfield  {booktitle} {\emph {\bibinfo {booktitle} {Reports on Progress in
  Physics}},\ }\href {\doibase 10.1088/1361-6633/ab37ca} {\bibfield  {journal}
  {\bibinfo  {journal} {Rep. Prog. Phys.}\ }\textbf {\bibinfo {volume} {82}},\
  \bibinfo {pages} {116601} (\bibinfo {year} {2019})}\BibitemShut {NoStop}%
\bibitem [{\citenamefont {Troutier}\ and\ \citenamefont
  {Ladavi{\`e}re}(2007)}]{Troutier:2007aa}%
  \BibitemOpen
  \bibfield  {author} {\bibinfo {author} {\bibfnamefont {A.-L.}\ \bibnamefont
  {Troutier}}\ and\ \bibinfo {author} {\bibfnamefont {C.}~\bibnamefont
  {Ladavi{\`e}re}},\ }\href {\doibase
  https://doi.org/10.1016/j.cis.2007.02.003} {\bibfield  {journal} {\bibinfo
  {journal} {Adv.\ Colloid Interf.\ Sci.}\ }\textbf {\bibinfo {volume} {133}},\
  \bibinfo {pages} {1} (\bibinfo {year} {2007})}\BibitemShut {NoStop}%
\bibitem [{\citenamefont {van~der Meulen}\ and\ \citenamefont
  {Leunissen}(2013)}]{VdMeulenJACS2013}%
  \BibitemOpen
  \bibfield  {author} {\bibinfo {author} {\bibfnamefont {S.~A.~J.}\
  \bibnamefont {van~der Meulen}}\ and\ \bibinfo {author} {\bibfnamefont
  {M.~E.}\ \bibnamefont {Leunissen}},\ }\href {\doibase 10.1021/ja406226b}
  {\bibfield  {journal} {\bibinfo  {journal} {J. Am. Chem. Soc.}\ }\textbf
  {\bibinfo {volume} {135}},\ \bibinfo {pages} {15129} (\bibinfo {year}
  {2013})},\ \Eprint
  {http://arxiv.org/abs/http://pubs.acs.org/doi/pdf/10.1021/ja406226b}
  {http://pubs.acs.org/doi/pdf/10.1021/ja406226b} \BibitemShut {NoStop}%
\bibitem [{\citenamefont {Zhang}\ and\ \citenamefont
  {Winfree}(2009)}]{zhang2009control}%
  \BibitemOpen
  \bibfield  {author} {\bibinfo {author} {\bibfnamefont {D.~Y.}\ \bibnamefont
  {Zhang}}\ and\ \bibinfo {author} {\bibfnamefont {E.}~\bibnamefont
  {Winfree}},\ }\href@noop {} {\bibfield  {journal} {\bibinfo  {journal} {J.
  Am. Chem. Soc.}\ }\textbf {\bibinfo {volume} {131}},\ \bibinfo {pages}
  {17303} (\bibinfo {year} {2009})}\BibitemShut {NoStop}%
\bibitem [{\citenamefont {Bomboi}\ \emph {et~al.}(2019)\citenamefont {Bomboi},
  \citenamefont {Caprara}, \citenamefont {Fernandez-Castanon},\ and\
  \citenamefont {Sciortino}}]{Bomboi2019}%
  \BibitemOpen
  \bibfield  {author} {\bibinfo {author} {\bibfnamefont {F.}~\bibnamefont
  {Bomboi}}, \bibinfo {author} {\bibfnamefont {D.}~\bibnamefont {Caprara}},
  \bibinfo {author} {\bibfnamefont {J.}~\bibnamefont {Fernandez-Castanon}}, \
  and\ \bibinfo {author} {\bibfnamefont {F.}~\bibnamefont {Sciortino}},\
  }\href@noop {} {\bibfield  {journal} {\bibinfo  {journal} {Nanoscale}\ ,\
  \bibinfo {pages} {9691}} (\bibinfo {year} {2019})}\BibitemShut {NoStop}%
\bibitem [{\citenamefont {Bachmann}\ \emph {et~al.}(2016)\citenamefont
  {Bachmann}, \citenamefont {Kotar}, \citenamefont {Parolini}, \citenamefont
  {{\v{S}}ari{\'c}}, \citenamefont {Cicuta}, \citenamefont {Di~Michele},\ and\
  \citenamefont {Mognetti}}]{BachmannSoftMatter2016}%
  \BibitemOpen
  \bibfield  {author} {\bibinfo {author} {\bibfnamefont {S.~J.}\ \bibnamefont
  {Bachmann}}, \bibinfo {author} {\bibfnamefont {J.}~\bibnamefont {Kotar}},
  \bibinfo {author} {\bibfnamefont {L.}~\bibnamefont {Parolini}}, \bibinfo
  {author} {\bibfnamefont {A.}~\bibnamefont {{\v{S}}ari{\'c}}}, \bibinfo
  {author} {\bibfnamefont {P.}~\bibnamefont {Cicuta}}, \bibinfo {author}
  {\bibfnamefont {L.}~\bibnamefont {Di~Michele}}, \ and\ \bibinfo {author}
  {\bibfnamefont {B.~M.}\ \bibnamefont {Mognetti}},\ }\href@noop {} {\bibfield
  {journal} {\bibinfo  {journal} {Soft Matter}\ }\textbf {\bibinfo {volume}
  {12}},\ \bibinfo {pages} {7804} (\bibinfo {year} {2016})}\BibitemShut
  {NoStop}%
\bibitem [{\citenamefont {Di~Michele}\ \emph {et~al.}(2016)\citenamefont
  {Di~Michele}, \citenamefont {Bachmann}, \citenamefont {Parolini},\ and\
  \citenamefont {Mognetti}}]{DiMicheleJCP2016}%
  \BibitemOpen
  \bibfield  {author} {\bibinfo {author} {\bibfnamefont {L.}~\bibnamefont
  {Di~Michele}}, \bibinfo {author} {\bibfnamefont {S.~J.}\ \bibnamefont
  {Bachmann}}, \bibinfo {author} {\bibfnamefont {L.}~\bibnamefont {Parolini}},
  \ and\ \bibinfo {author} {\bibfnamefont {B.~M.}\ \bibnamefont {Mognetti}},\
  }\href {\doibase 10.1063/1.4947550} {\bibfield  {journal} {\bibinfo
  {journal} {J. Chem. Phys.}\ }\textbf {\bibinfo {volume} {144}},\ \bibinfo
  {pages} {161104} (\bibinfo {year} {2016})},\ \Eprint
  {http://arxiv.org/abs/https://doi.org/10.1063/1.4947550}
  {https://doi.org/10.1063/1.4947550} \BibitemShut {NoStop}%
\bibitem [{\citenamefont {Sciortino}\ \emph {et~al.}(2020)\citenamefont
  {Sciortino}, \citenamefont {Zhang}, \citenamefont {Gang},\ and\ \citenamefont
  {Kumar}}]{Sciortino2020}%
  \BibitemOpen
  \bibfield  {author} {\bibinfo {author} {\bibfnamefont {F.}~\bibnamefont
  {Sciortino}}, \bibinfo {author} {\bibfnamefont {Y.}~\bibnamefont {Zhang}},
  \bibinfo {author} {\bibfnamefont {O.}~\bibnamefont {Gang}}, \ and\ \bibinfo
  {author} {\bibfnamefont {S.~K.}\ \bibnamefont {Kumar}},\ }\href@noop {}
  {\bibfield  {journal} {\bibinfo  {journal} {ACS Nano}\ }\textbf {\bibinfo
  {volume} {5}},\ \bibinfo {pages} {5628} (\bibinfo {year} {2020})}\BibitemShut
  {NoStop}%
\bibitem [{\citenamefont {Cerbino}\ and\ \citenamefont
  {Trappe}(2008)}]{Cerbino:2008aa}%
  \BibitemOpen
  \bibfield  {author} {\bibinfo {author} {\bibfnamefont {R.}~\bibnamefont
  {Cerbino}}\ and\ \bibinfo {author} {\bibfnamefont {V.}~\bibnamefont
  {Trappe}},\ }\href {\doibase 10.1103/PhysRevLett.100.188102} {\bibfield
  {journal} {\bibinfo  {journal} {Phys. Rev. Lett.}\ }\textbf {\bibinfo
  {volume} {100}},\ \bibinfo {pages} {188102} (\bibinfo {year}
  {2008})}\BibitemShut {NoStop}%
\bibitem [{\citenamefont {Cerbino}\ and\ \citenamefont
  {Cicuta}(2017)}]{Cerbino:2017aa}%
  \BibitemOpen
  \bibfield  {author} {\bibinfo {author} {\bibfnamefont {R.}~\bibnamefont
  {Cerbino}}\ and\ \bibinfo {author} {\bibfnamefont {P.}~\bibnamefont
  {Cicuta}},\ }\bibfield  {booktitle} {\emph {\bibinfo {booktitle} {The Journal
  of Chemical Physics}},\ }\href {\doibase 10.1063/1.5001027} {\bibfield
  {journal} {\bibinfo  {journal} {J. Chem. Phys.}\ }\textbf {\bibinfo {volume}
  {147}},\ \bibinfo {pages} {110901} (\bibinfo {year} {2017})}\BibitemShut
  {NoStop}%
\bibitem [{\citenamefont {Cho}\ \emph {et~al.}(2020)\citenamefont {Cho},
  \citenamefont {Cerbino},\ and\ \citenamefont {Bischofberger}}]{Cho:2020aa}%
  \BibitemOpen
  \bibfield  {author} {\bibinfo {author} {\bibfnamefont {J.~H.}\ \bibnamefont
  {Cho}}, \bibinfo {author} {\bibfnamefont {R.}~\bibnamefont {Cerbino}}, \ and\
  \bibinfo {author} {\bibfnamefont {I.}~\bibnamefont {Bischofberger}},\ }\href
  {\doibase 10.1103/PhysRevLett.124.088005} {\bibfield  {journal} {\bibinfo
  {journal} {Phys. Rev. Lett.}\ }\textbf {\bibinfo {volume} {124}},\ \bibinfo
  {pages} {088005} (\bibinfo {year} {2020})}\BibitemShut {NoStop}%
\bibitem [{\citenamefont {Lanfranco}\ \emph {et~al.}(2019)\citenamefont
  {Lanfranco}, \citenamefont {Jana}, \citenamefont {Tunesi}, \citenamefont
  {Cicuta}, \citenamefont {Mognetti}, \citenamefont {Di~Michele},\ and\
  \citenamefont {Bruylants}}]{LanfrancoLangmuir2019}%
  \BibitemOpen
  \bibfield  {author} {\bibinfo {author} {\bibfnamefont {R.}~\bibnamefont
  {Lanfranco}}, \bibinfo {author} {\bibfnamefont {P.~K.}\ \bibnamefont {Jana}},
  \bibinfo {author} {\bibfnamefont {L.}~\bibnamefont {Tunesi}}, \bibinfo
  {author} {\bibfnamefont {P.}~\bibnamefont {Cicuta}}, \bibinfo {author}
  {\bibfnamefont {B.~M.}\ \bibnamefont {Mognetti}}, \bibinfo {author}
  {\bibfnamefont {L.}~\bibnamefont {Di~Michele}}, \ and\ \bibinfo {author}
  {\bibfnamefont {G.}~\bibnamefont {Bruylants}},\ }\href {\doibase
  10.1021/acs.langmuir.8b02707} {\bibfield  {journal} {\bibinfo  {journal}
  {Langmuir}\ }\textbf {\bibinfo {volume} {35}},\ \bibinfo {pages} {2002}
  (\bibinfo {year} {2019})},\ \Eprint
  {http://arxiv.org/abs/https://doi.org/10.1021/acs.langmuir.8b02707}
  {https://doi.org/10.1021/acs.langmuir.8b02707} \BibitemShut {NoStop}%
\bibitem [{\citenamefont {Buddingh'}\ and\ \citenamefont {van
  Hest}(2017)}]{Buddingh:2017aa}%
  \BibitemOpen
  \bibfield  {author} {\bibinfo {author} {\bibfnamefont {B.~C.}\ \bibnamefont
  {Buddingh'}}\ and\ \bibinfo {author} {\bibfnamefont {J.~C.~M.}\ \bibnamefont
  {van Hest}},\ }\bibfield  {booktitle} {\emph {\bibinfo {booktitle} {Accounts
  of Chemical Research}},\ }\href {\doibase 10.1021/acs.accounts.6b00512}
  {\bibfield  {journal} {\bibinfo  {journal} {Acc. Chem. Res.}\ }\textbf
  {\bibinfo {volume} {50}},\ \bibinfo {pages} {769} (\bibinfo {year}
  {2017})}\BibitemShut {NoStop}%
\bibitem [{\citenamefont {van~der Vaart}\ and\ \citenamefont
  {Pretorius}(2007)}]{Vaart:2007aa}%
  \BibitemOpen
  \bibfield  {author} {\bibinfo {author} {\bibfnamefont {M.}~\bibnamefont
  {van~der Vaart}}\ and\ \bibinfo {author} {\bibfnamefont {P.~J.}\ \bibnamefont
  {Pretorius}},\ }\href {\doibase 10.1373/clinchem.2007.092734} {\bibfield
  {journal} {\bibinfo  {journal} {Clinical Chem.}\ }\textbf {\bibinfo {volume}
  {53}},\ \bibinfo {pages} {2215} (\bibinfo {year} {2007})}\BibitemShut
  {NoStop}%
\bibitem [{\citenamefont {Li}\ \emph {et~al.}(2009)\citenamefont {Li},
  \citenamefont {Raoult},\ and\ \citenamefont {Fournier}}]{Li:2009aa}%
  \BibitemOpen
  \bibfield  {author} {\bibinfo {author} {\bibfnamefont {W.}~\bibnamefont
  {Li}}, \bibinfo {author} {\bibfnamefont {D.}~\bibnamefont {Raoult}}, \ and\
  \bibinfo {author} {\bibfnamefont {P.-E.}\ \bibnamefont {Fournier}},\ }\href
  {\doibase 10.1111/j.1574-6976.2009.00182.x} {\bibfield  {journal} {\bibinfo
  {journal} {FEMS Microbiol. Rev.}\ }\textbf {\bibinfo {volume} {33}},\
  \bibinfo {pages} {892} (\bibinfo {year} {2009})}\BibitemShut {NoStop}%
\bibitem [{\citenamefont {Mackay}\ \emph {et~al.}(2002)\citenamefont {Mackay},
  \citenamefont {Arden},\ and\ \citenamefont {Nitsche}}]{Mackay:2002aa}%
  \BibitemOpen
  \bibfield  {author} {\bibinfo {author} {\bibfnamefont {I.~M.}\ \bibnamefont
  {Mackay}}, \bibinfo {author} {\bibfnamefont {K.~E.}\ \bibnamefont {Arden}}, \
  and\ \bibinfo {author} {\bibfnamefont {A.}~\bibnamefont {Nitsche}},\ }\href
  {\doibase 10.1093/nar/30.6.1292} {\bibfield  {journal} {\bibinfo  {journal}
  {Nucl. Acids Res.}\ }\textbf {\bibinfo {volume} {30}},\ \bibinfo {pages}
  {1292} (\bibinfo {year} {2002})}\BibitemShut {NoStop}%
\bibitem [{\citenamefont {O'Brien}\ \emph {et~al.}(2018)\citenamefont
  {O'Brien}, \citenamefont {Hayder}, \citenamefont {Zayed},\ and\ \citenamefont
  {Peng}}]{OBrien:2018aa}%
  \BibitemOpen
  \bibfield  {author} {\bibinfo {author} {\bibfnamefont {J.}~\bibnamefont
  {O'Brien}}, \bibinfo {author} {\bibfnamefont {H.}~\bibnamefont {Hayder}},
  \bibinfo {author} {\bibfnamefont {Y.}~\bibnamefont {Zayed}}, \ and\ \bibinfo
  {author} {\bibfnamefont {C.}~\bibnamefont {Peng}},\ }\href
  {https://www.frontiersin.org/article/10.3389/fendo.2018.00402} {\bibfield
  {journal} {\bibinfo  {journal} {Front. Endocrin.}\ }\textbf {\bibinfo
  {volume} {9}},\ \bibinfo {pages} {402} (\bibinfo {year} {2018})}\BibitemShut
  {NoStop}%
\end{thebibliography}

%

\end{document}



\title{Electronic Supplementary Information:\\ {  Adaptable}  DNA  interactions  regulate  surface triggered self assembly .}

\author{Roberta Lanfranco}
 \affiliation{
 Cavendish Laboratory, University of Cambridge, JJ Thomson Avenue, Cambridge, CB3 0HE, United Kingdom
 }
 
\author{Pritam Kumar Jana}%
\affiliation{
Interdisciplinary Center for Nonlinear Phenomena and Complex Systems, Universit\'{e} libre de Bruxelles (ULB) Campus Plaine, CP 231, Blvd. du Triomphe, B-1050 Brussels, Belgium
}

\author{Gilles Bruylants}
\affiliation{
Engineering of Molecular NanoSystems, Universit\'{e} libre de Bruxelles (ULB), 50 av. F.D. Roosevelt, 1050 Brussels, Belgium
}

\author{Pietro Cicuta}
 \affiliation{
 Cavendish Laboratory, University of Cambridge, JJ Thomson Avenue, Cambridge, CB3 0HE, United Kingdom
 }
 
\author{Bortolo Matteo Mognetti}
\affiliation{
Interdisciplinary Center for Nonlinear Phenomena and Complex Systems, Universit\'{e} libre de Bruxelles (ULB) Campus Plaine, CP 231, Blvd. du Triomphe, B-1050 Brussels, Belgium
}

\author{Lorenzo Di Michele}
  \email{l.di-michele@imperial.ac.uk}
  \affiliation{
  Department of Chemistry, Imperial College London, Molecular Sciences Research Hub, 80 Wood Lane, London W12 0BZ, United Kingdom}
 \affiliation{
 Cavendish Laboratory, University of Cambridge, JJ Thomson Avenue, Cambridge, CB3 0HE, United Kingdom
 }

\date{\today}

\pacs{Valid PACS appear here}
\maketitle
\section{Experimental Methods}

\subsection{Sample Preparation}

\subsubsection{Preparation of supported lipid bilayers (SLBs) on particles and substrate spheres}
The protocol to coat silica particles ($\diameter = 0.985\pm0.04$\,$\muup$m, Microparticles GmbH) and substate spheres ($\diameter = 9.56\pm0.25$\,$\muup$m  Microparticles GmbH) with a SLB was adapted from Ref.~\citen{Rinaldin:2019aa}. We first prepared, in a glass vial, a chloroform solution of 98\% molar fraction DOPC (1,2-dioleoyl-sn-glycero-3-phosphocholine, Avanti Polar Lipids),  1\% molar fraction DHPE--Texas Red (Texas Red 1,2-Dihexadecanoyl-sn-Glycero-3-Phosphoethanolamine, Triethylammonium Salt, Invitrogen), and 1\% molar fraction of PEG(2000)-DOPE (1,2-dioleoyl-sn-glycero-3-phosphoethanolamine-N-[methoxy(polyethylene glycol)-2000, Avanti Polar Lipids).  The fluorescently-tagged lipid was used for visualising the objects in confocal experiments, while the PEGylated lipids prevent non specific aggregation during functionalization steps. For FRAP experiments aimed at assessing the mobility of (fluorescent) anchored DNA constructs the fluorescent lipid was not used and replaced with DOPC. The lipid solution was dried under vacuum for 20 minutes and left in a desiccator overnight to form a dry lipd film, which was then re-hydrated in a low ionic-strength buffer (50 mM NaCl + 1$\times$ TE buffer + 0.1\%  w/v NaN$_3$,  pH 7.4) to obtain a total lipid concentration of 1\,mg\,ml$^{-1}$. Small liposomes were then produced using a tip sonicator (cycle of 300 ms, 30\% power for 20 minutes). To remove the particulate left by the tip, the liposome sample was centrifuged for 1h at 17000 rcf, and the liposome-containing supernatant collected for the next step.\\
 The liposome solution was then mixed with silica particles and spheres with an estimated $10\times$ excess of lipid bilayer compared with the overall area of the silica particles/spheres. The sample were left under gentle agitation for at least 3 hours, to promote the formation of the SLB.  Afterwards, the sample was diluted with a buffer with no added salt (1$\times$ TE buffer + 0.1\% w/v NaN$_3$,  pH 7.4), reducing the NaCl concentration to 12.5 mM. To remove the lipid excess, particles were made to sediment by gentle centrifugation (4 minutes at 1200 rcf), while 10 micron particles were left to sediment naturally for for 15 minutes as centrifugation was found to substantially damage the SLB. The supernatant was finally replaced with 1$\times$ TE buffer + 0.1\% w/v NaN$_3$ (pH 7.4, no added salt) and this procedure was repeated for 5 times. This protocol allowed for the formation of a continuous bilayer around the small particles. Discontinuous (patchy) SLB were instead formed on a fraction of the substrate spheres. These could simply be disregarded when analysing the data on layer formation, having demonstrated that the presence of substrate spheres has no effect on the bulk phase behaviour of the particles (see Fig.~\ref{FigureS3}).\\
 
 \subsubsection{Preparation DNA linkers, inert constructs and fluorescent DNA probes}
Linkers and other DNA constructs were prepared from individual single-stranded DNA components, the sequences of which are reported in Table~\ref{table}. All constructs featured two ssDNA strands labelled with cholesterol/cholesteryl which form a 18 bp duplex with a 18 nt overhang.  Two different cholesterolised DNA duplexes were used in this work, formed from strands $CHA_{1}+CHA_{2}$ and $CHB_{1}+CHB_{2}$, respectively. To create linkers, sticky end sequences $SEA_1$ $SEA_2$ bind to the overhangs of cholesterolised duplex $CHA_{1}+CHA_{2}$, while $SEB$ and $SEC$ bind to $CHB_{1}+CHB_{2}$. Four unpaired Thymines were left between the spacer of the formed linker and  the sticky end, to enable accessibility of the domains and flexibility. Inert constructs were prepared from ssDNA strands $I_1$ and $I_2$, forming a 32 bp duplex with a 18 bp overhang fully complementary to that of $CHB_{1}+CHB_{2}$ cholestrolised duplex. For fluorescent DNA probes used in FRAP experiments (Fig.~\ref{FigureS1}) cholesterolised duplexes $CHA_{1}+CHA_{2}$ and $CHB_{1}+CHB_{2}$ were coupled to labelled oligos $Fluo_1$ and $Fluo_2$, respectively.\\
Each construct type was individually prepared by mixing all the single-stranded components in stoichiometric ratio at a concentration of $10$\,$\muup$M in TE buffer + 100 mM NaCl. Samples were then heated up to 96$^\circ$C and let cool down to 20$^\circ$C over 4 hours on a thermal cycler to favour self-assembly.\\

\subsubsection{Functionalisation of SLB-coated particles and substrates with  DNA constructs}
To enable insertion of cholesterolised DNA constructs in the membranes surrounding the silica particles and spheres, the latter were combined with suitable mixtures of constructs. The salt concentration of the TE buffer solution was adjusted to 50 mM NaCl. For particles, the concentration of different linker types was chosen such that $[A_1]=[A_2]=0.5\times[B]$, and $[L]/([L]+[I])=f=0.0, 0.1, 0.2, 0.3, 0.4$, where $[L]=[A_1]+[A_2]+[B]$. The overall concentration of constructs was fixed to achieve a nominal total number of constructs per particle equal to 1.6$\times10^5$. For substrate particles, a $\sim20\%$ excess of linkers $C$ was added in solution to guarantee the highest possible coverage.\\
After 15 hours, possible DNA constructs remaining in solution were removed by sedimentation and supernatant exchange, repeated 5 times. As done for removal of lipid excess, sedimentation was induced by gentle centrifugation for the particles and occurs naturally for the substrate spheres. The buffer used for the washing steps is the final experimental buffer (100 mM NaCl + 1$\times$ TE buffer + 0.1\% w/v NaN$_3$,  pH 7.4). To aid resuspension and break possible non-specific clumps, samples were sonicated for 30\,s between each washing step.\\
Before microscopy experiments, particles carrying active DNA strands were heated-up to 60$^\circ$C for 10 minutes and then the temperature was rapidly quenched to 10$^\circ$C to favour the formation of loops instead of bridges, a procedure previously applied to liposomes functionalised with similar constructs~\cite{parolini2016controlling}. 

\subsubsection{Preparation of microscopy chambers\label{chamebrs}} 
Borosilicate glass coverslips (20\,mm$\times$40\,mm no.1, Menzel) were cleaned by sonicating four times for 15 minutes. The first sonication step was performed in  1\% (volume) Hellmanex solution (Hellma), the second in ultrapure water, the third in 96\% Ethanol, and the fourth in ultrapure water. Slides were thoroughly rinsed with ultrapure water between each step.\\
Clean and dry particles were then silanised, by placing them in a dessicator with a few droplets of 1H,1H,2H,2H-Perfluorodecyltrichlorosilane (96\%, Thermo Fisher). The dessicator was placed under vacuum for 10 minutes, and then left overnight.\\
Sticky silicone rubber chambers (FlexWells  incubation chambers, Garce Biolabs) were then applied to the silanised coverslips to form wells. Chambers were passivated with block co-polymer Pluronic F-127 (Sigma) by filling them with a 0.1\%  w/v solution in experimental buffer  (100 mM NaCl + 1$\times$ TE buffer + 0.1\% w/v NaN$_3$,  pH 7.4) and incubating for 30 minutes. { Passivation was required to prevent non-specific adhesion of particles to the chamber bottom and their consequent immobilisation.} Finally the chambers were rinsed in experimental buffer and filled with relevant particles and substrates. A small concentration of Pluronic  (0.05\% wt) in the final experimental buffer was included to prevent non-specific adhesion of the particles to the glass bottom of the chamber. The composition of the experimental buffer used for microscopy experiments is therefore  100 mM NaCl + 1$\times$ TE buffer + 0.1\% w/v NaN$_3$ + 0.05\% w/v Pluronic F-127, pH 7.4. { The small amount of free Pluronic F-127 was included to prevent polymer desorption over the course of the experiments.}\\
For all samples, an overall particle concentration of {0.12\%} w/v was used. Note however that silica particles have a barometric height of roughly 3\,$\muup$m, so our system can be regarded as quasi-2D, with an effective packing fraction in the bottom 10\,$\muup$m of the chamber of $\sim 3-5\%$,  as determined from image analysis. A small number of substate spheres (30 to 40) was present in each well. Substrate spheres sediment readily and do not display height thermal fluctuations.\\

\subsection{Imaging and data analysis}

\subsubsection{Differential Dynamic Microscopy}~\label{DDM}
For DDM experiments~\cite{Cerbino:2008aa,Cerbino:2017aa}, samples were imaged with a fully automated Nikon Ti-E inverted microscope equipped with Perfect Focusing System. Imaging was done in bright field mode using a Nikon CFI Plan APO 20$\times$ 0.75 NA dry objective and a Ximea camera. We collected 20-second videos at 50 fps, at 1 hour intervals for 22 hours. Two locations in each sample were imaged, and each video was further divided in four regions of interest (ROIs). Videos from each field of view and ROI were processed separately using a tailor made script for DDM to extract the image structure function (Eq. 3 in Ref. ~\citen{Cerbino:2008aa}) and the decay times $\tau(q)$ corresponding to the Fourier modes of wave vector $q$. Examples of  $\tau(q)$ measured for samples with different fraction of linkers $f$ are shown in Fig.~\ref{FigureS3}.  The  curves were fitted as $\tau(q)=Dq^{-2}$ to extract an effective diffusion coefficient $D$. Note that, as demonstrated in Fig.~\ref{FigureS3}, $\tau(q)$ curves are best fitted with a power law $\propto q^{-\alpha}$, with $\alpha<2$. The deviation from the ideal Brownian behaviour ($\alpha=2$) is particularly prominent for samples with substantial particle aggregation, \emph{e.g.} for large $f$ and at late experimental stages, and is ascribed to the dynamic heterogeneity of the colloidal clusters and gels~\cite{Cho:2020aa}. Nonetheless, diffusion coefficient extracted from the Brownian fit was used to assess the presence of particle aggregation in Fig.~2\textbf{b} and Fig.~\ref{FigureS3}, as it still represents a good indicator of the aggregation state of the sample.

\subsubsection{Confocal Imaging}
To assess the number and arrangement of particles adhering to substate spheres we performed confocal imaging on a Leica SP5II point-scanning confocal microscope, equipped with a HCX PL Fluotar 63$\times$ 1.25 NA oil immersion objective. { Imaging was carried out $\sim$ 24 hours after sample preparation, to enable equilibration of the surface-triggered aggregates and having already characterised the presence (or absence) or bulk aggregation with time-lapse DDM experiments (Section~\ref{DDM}, Fig.~\ref{FigureS3}).} To image the Texas Red-tagged lipids on the SLBs we excited with a HeNe laser (596\,nm). We collected zoomed-in z-stacks of a large number of individual substrate spheres. Stacks were recorded in both confocal (centre in Fig.~3) and transmission bright field mode  (top in Fig. 3). Individual z-stacks were processed with a tailor-made Matlab script to track the location of adhering particles and determine the ``layer'' they belong to, obtaining the histograms in Fig 3 (bottom). The script operates as follows:
\begin{itemize}
\item A z-stack (with both confocal and bright-field frames) featuring a substrate sphere is randomly selected from a folder containing data for all $f$ values, blinding the analysis and avoiding human bias in the manual steps (see below).
\item The 3D coordinates ($x_\mathrm{s}$, $y_\mathrm{s}$, $z_\mathrm{s}$) of the centre of the substrate particle and its radius ($R$) are detected from the bright field data using a circle-finding routine. The value of $R$ is then checked and, if needed, refined by manual selection on the confocal images. The correction is performed manually as the small adhering particles makes automated detection of the large sphere challenging in confocal frames. 
\item The 3D coordinates of the particles are determined from confocal data. The z-coordinates ($z_{i}$) are determined manually by identifying the z-slice in which the particles are best in focus. The accuracy is limited by the separation of the z-slices ($0.1~\mu$\,m) , but the associated uncertainty is deemed negligible compared to other localisation errors. At this stage, particles which are not adhering to the substrate spheres are observed to quickly diffuse between subsequent frames of the z-stack, and are excluded from the analysis. The horizontal coordinates ($x_{i}$, $y_{i}$) and radii ($r_{i}$) are then determined by automated localisation on the relevant z-plane. 
\item The average particle radius $(r)$, used in the following analysis, is determined as the mean over all $r_{i}$.
\item Layers in Fig.~3 are defined as spherical shells around the centre of the substrate sphere. The first layer spans the distance interval $(R, R+r]$, while the $j$th layer spans the interval $(R_j,R_{j+1}]$, with $R_j=R+(2j-3)r$ and $j=2,3\dots$.
\item The distances $d_i$ between the centre of each particle and that of the substrate sphere is calculated, and the particle assigned to one layer based on the definition above.
\end{itemize}
For each $f$-value we imaged between 5 and 10 substrate spheres.\\

\subsubsection{FRAP measurements}
\begin{figure}[ht!]
\centering
  \includegraphics[width=17cm]{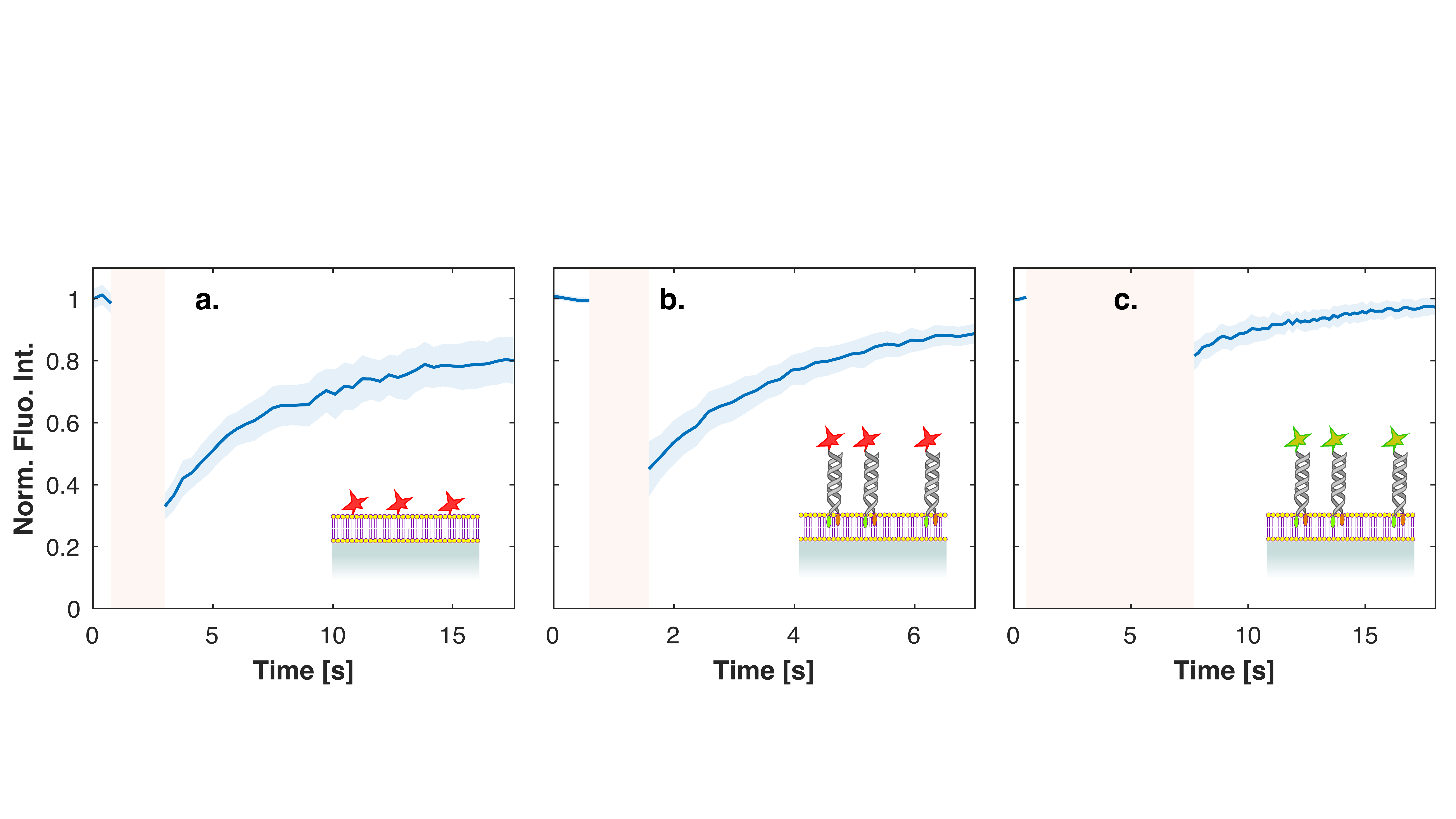}
  \caption{\textbf{FRAP experiments on substrate spheres}. FRAP recovery curves as recorded on SLB-coated substrate spheres probing DHPE-TexasRed lipids (\textbf{a}), Cy5-functionalised DNA constructs (\textbf{b}) and Cy3-functionalised DNA constructs (\textbf{c}). Spheres in \textbf{a} were also decorated with non-fluorescent inert DNA constructs to accurately represent the experimental scenario. Spheres used for \textbf{b} and \textbf{c} lack the fluorescent lipids in their SLB. Sequences of the ssDNA components of the constructs used in \textbf{b} and \textbf{c}, which differ for the cholesterolised membrane-anchoring element, are summarised in Table~\ref{table}. The shaded regions in all plots indicates the bleaching period, and its duration changes from sample to sample due to differences in the intensity of the relevant laser lines and the tendency to bleach of the different dyes. Curves are averaged over $\geq 6$ independent measurements performed on different spheres. The solid line and the shaded region surrounding it represent the mean and standard deviation of these measurements. In all cases, a clear recovery of the fluorescence is observed, demonstrating the lateral mobility of the tested probes. The timescales of the recovery are comparable with literature values for SLB on silica particles.~\cite{Rinaldin:2019aa}}
  \label{FigureS1}
\end{figure}

FRAP on the substrate spheres was performed on the Leica SP5 II confocal using the same objective described above, and taking advantage of the Leica FRAP wizard, to assess the mobility of lipids in the SLB and the anchored DNA constructs. Two of the latter were tested, one featuring the $CHA_1$+$CHA_2$ cholesterolised duplex (Cy5-functionalsied \emph{via} the $Fluo_1$ strand) and the second using the  $CHB_1$+$CHB_2$ cholesterolised duplex (Cy3-functionalsied  \emph{via} the $Fluo_2$ strand, see Table~\ref{table}). Bleaching and imaging were carried out with the 596\,nm HeNe laser when testing the diffusivity of the Texas Red tagged lipids (Fig.~\ref{FigureS1}\textbf{a}), a 633\,nm HeNe laser when testing Cy5-labelled DNA constructs (Fig.~\ref{FigureS1}\textbf{b}), and a 514\,nm Ar-ion line when testing Cy3-labelled DNA constructs (Fig.~\ref{FigureS1}\textbf{c}). Data were analysed with ImageJ by measuring the average pixel intensity within the bleached ROI and normalising it by the pre-bleach value. Data were also corrected for the effect of imaging-induced photobleaching by normalising for the fluorescence recorded on the substrate spheres outside the bleached spot. Due to their small size, FRAP experiments could not be reliably performed on the small particles.

\clearpage

\section{Theoretical and Numerical Methods}

\subsection{Multivalent Free-Energy}
We consider particles functionalized by three types of linkers ($A_1$,  $A_2$, $B$, and $I$, see Fig.~\ref{Fig:complexes}). $A_1$ and  $A_2$ can bind (also simultaneously)  $B$, while $I$ is an inert linker used to modulate the repulsive part of the interaction. $N_{A_1}$, $N_{A_2}$, $N_{B}$, and $N_{I}$ are the number of different linkers found on each particle. The partition function of a system with $N_\mr{p}$ particles is
\begin{align}
\label{Eq:partitionfunction}
	\begin{split}
	Z &= \frac{1}{N_\mr{p}!}\int d\{\mbf{r}\}\sum_{\{n\}}\mathcal{Z}(\{ n\},\{\mbf{r}\}) e^{-\beta F_\mathrm{rep}(\{ {\bf r} \})}
	\\
	&=\frac{1}{N_\mr{p}!}\int d\{\mbf{r}\}\sum_{\{n\}}e^{-\beta\mathcal{F}_\mr{multi}(\{ n\},\{\mbf{r}\})-\beta F_\mathrm{rep}(\{ {\bf r} \})},
		\end{split}
\end{align}
where $\{\mbf{r}\}$ is the list of the cartesian coordinates of the particles and $\{ n\}$ is the ensemble of possible inter-particle and intra-particle complexes. Fig.~\ref{Fig:complexes} reports some examples of inter-particle and intra-particle complexes (the full list is detailed in Eqs.~\ref{Eq:intra-particle}, \ref{Eq:inter-particle}). $n_{i}^{A_1}$, $n_{i}^{A_2}$, and $n_{j}^{B}$ are the number of free linkers on particle $i$. $\mathcal{Z}$ and $\mathcal{F}_\mr{multi}$ represent the partition function of the system and the multivalent free energy, respectively, at a given $\{\mbf{r}\}$ and $\{ n\}$. $F_\mathrm{rep}$ accounts for non-specific interactions and repulsive terms detailed in the next section. 
$\mathcal{Z}$ comprises combinatorial terms, counting the number of ways of making a given set of complexes $\{ n \}$, and hybridization free energies ($\Delta G_{0}^{BA_1}$, $\Delta G_{0}^{BA_2}$, and $\Delta G_{0}^{BA_1A_2}$).
At a given colloid position $\{\mbf{r}\}$, the most likely numbers of bonds featured by the system are obtained by minimising the multivalent free energy $\mathcal{F}_\mr{multi}$ \cite{mognetti2019programmable}
\begin{equation}
\frac{\partial}{\partial{n}}\mathcal{F}_\mr{multi}(\{n\})=0
\label{Eq:freeenergyminimization}
\end{equation}
Generally, Eqs.~\ref{Eq:freeenergyminimization} are equivalent to chemical equilibrium equations for the different types of complexes \cite{mognetti2019programmable}. For intra-particle loops we have
\begin{align}
	\label{Eq:intra-particle}
	\begin{split}
	{n}_{ii}^{BA_1} &={n}_{i}^{{B}}{n}_{i}^{B_1}q_l\exp[-\beta\Delta G_{0}^{BA_1}]
	\\
	{n}_{ii}^{BA_2} &={n}_{i}^{{B}}{n}_{i}^{B_2}q_l\exp[-\beta\Delta G_{0}^{BA_2}]
	\\
	{n}_{ii}^{BA_1A_2} &={n}_{i}^{{B}}{n}_{i}^{B_1}{n}_{i}^{B_2}q_lq_l\exp[-\beta\Delta G_{0}^{BA_1A_2}]
		\end{split}
\end{align}
while for inter-particle bridges  
\begin{align}
	\label{Eq:inter-particle}
	\begin{split}
		{n}_{ij}^{B;A_1} &={n}_{i}^{{B}}{n}_{j}^{B_1}q_b\exp[-\beta\Delta G_{0}^{BA_1}]
		\\
		{n}_{ij}^{B;A_2} &={n}_{i}^{{B}}{n}_{j}^{B_2}q_b\exp[-\beta\Delta G_{0}^{BA_2}]
		\\
		{n}_{ij}^{B_1;B} &={n}_{i}^{{B_1}}{n}_{j}^{B}q_b\exp[-\beta\Delta G_{0}^{B_1B}]
		\\
		{n}_{ij}^{B_2;B} &={n}_{i}^{{B_2}}{n}_{j}^{B}q_b\exp[-\beta\Delta G_{0}^{B_2B}]
		\\
		{n}_{ij}^{B_1;A_2B} &={n}_{i}^{{B_1}}{n}_{j}^{B_2}{n}_{j}^{B}q_lq_b\exp[-\beta\Delta G_{0}^{BA_1A_2}]
		\\
		{n}_{ij}^{B_2;A_1B} &={n}_{i}^{{B_2}}{n}_{j}^{B_1}{n}_{j}^{B}q_lq_b\exp[-\beta\Delta G_{0}^{BA_1A_2}]
		\\
		{n}_{ij}^{B;A_1A_2} &={n}_{i}^{{B}}{n}_{j}^{B_1}{n}_{j}^{B_2}q_lq_b\exp[-\beta\Delta G_{0}^{BA_1A_2}]
		\\
		{n}_{ij}^{B_1A_2;B} &={n}_{i}^{{B_1}}{n}_{i}^{{B_2}}{n}_{j}^{B}q_lq_b\exp[-\beta\Delta G_{0}^{BA_1A_2}]
		\\
		{n}_{ij}^{B_1B;A_2} &={n}_{i}^{{B_1}}{n}_{i}^{{B}}{n}_{j}^{{B_2}}q_lq_b\exp[-\beta\Delta G_{0}^{BA_1A_2}]
		\\
		{n}_{ij}^{B_2B;A_1} &={n}_{i}^{{B_2}}{n}_{i}^{{B}}{n}_{j}^{{B_1}}q_lq_b\exp[-\beta\Delta G_{0}^{BA_1A_2}]
	\end{split}
\end{align}
with $\beta=(k_BT)^{-1}$.
 $\Delta G_{0}^{BA_1}$, $\Delta G_{0}^{BA_2}$, $\Delta G_{0}^{BA_1A_2}$ are the hybridization free energies of forming ${B_1}{B}$, ${B_2}{B}$, and ${BA_1A_2}$ complexes starting from free linkers in solution using as reference concentration $\rho_0$, $\rho_0=1\,$mol/litre. If $\Delta G_0=\Delta H_0 - T \Delta S_0$, in this study \cite{markham2005dinamelt,parolini2016controlling}
 \begin{eqnarray}
 \Delta H_{0}^{BA_1} = -63.3 \,\mathrm{Kcal/mol} & \qquad &  
  \Delta S_{0}^{BA_1} = -177.1 \,\mathrm{cal/mol/K}
 \\
 \Delta H_{0}^{BA_2} =-58.6\,\mathrm{Kcal/mol}  & \qquad &
  \Delta S_{0}^{BA_2} =-161.5\,\mathrm{cal/mol/K}
 \\
 \Delta H_{0}^{BA_1A_2} =-85.2\,\mathrm{Kcal/mol} & \qquad &
 \Delta S_{0}^{BA_1A_2} =-241.4\,\mathrm{cal/mol/K}
 \end{eqnarray}

Linkage formation leads to a loss of configurational entropy, which is denoted as  $q_b$ for bridge formation and $q_l$ for loop formation.  In particular \cite{mognetti2019programmable} 
\begin{align}
	\label{Eq:confcost}
	\begin{split}
		q_b = \frac{\Omega_{ij}(\{\mbf{r}\})}{\Omega_i(\{\mbf{r}\})\Omega_j(\{\mbf{r}\})\rho_0}
		\\
		q_l = \frac{1}{\Omega_i(\{\mbf{r}\})\rho_0}
	\end{split}
\end{align}
where $\Omega_{ij}$ is the volume available to the reacted sticky ends (assumed point-like) of bridges made of linkers tethered to $i$ and $j$ (see \ref{Fig:overlapvolume}) and $\Omega_{i}$ the volume available to the reactive sticky ends of free linkers.  Defining $e_{ij}$ as the volume excluded to the free linkers tethered to $i$ by the presence of particle $j$ (see Fig.~\ref{Fig:overlapvolume}) we have 
\begin{eqnarray}
\Omega_i = \Omega_0 - \sum_{j\in \langle i \rangle} e_{ij} 
\end{eqnarray}
where $\langle i \rangle$ is the list of particles interacting with $i$ and $\Omega_0=4\pi R^2 L$.
 The expressions of $\Omega_{ij}$ and $e_{ij}$ follow
\begin{align}
	\label{Eq:overlapvolume}
	\begin{split}
		\Omega_{ij}(r_{ij},L)=v(r_{ij}, R+L, R+L)-2v(r_{ij},R,R)
	\end{split}
\end{align}
\begin{align}
	\label{Eq:depletedvolume}
	\begin{split}
			e_{ij}(r_{ij},L)=v(r_{ij}, R+L, R),
	\end{split}
\end{align}
where $v(r, R_1, R_2)$ is the overlapping volume between two spheres of radius $R_1$ and $R_2$ placed at a distance $r$,  
\begin{align}
	\label{Eq:overlapvolume1}
	\begin{split}
		v(r,R_1,R_2)=\frac{\pi}{12r}(R_1+R_2-r)^2(r^2+2rR_1+2rR_2-3R_1^2-3R_2^2+6R_1R_2).
	\end{split}
\end{align}

Using the solutions of Eqs.~\ref{Eq:intra-particle}, \ref{Eq:inter-particle} into ${\cal F}_\mr{multi}$ (Eq.~\ref{Eq:partitionfunction}) one obtains the following portable expression of the multivalent free-energy

\begin{equation}
\label{Eq:FreeEn}
\begin{aligned}
\beta F_\mr{multi}(\{\mathbf{r}\}) = &\sum_{i=1}^{N_\mr{p}} \left(N_{A_1}\log\frac{{n}_{i}^{{B_1}}}{N_{A_1}}+N_{A_2}\log\frac{{n}_{i}^{{B_2}}}{N_{A_2}}+N_{B}\log\frac{{n}_{i}^{{B}}}{N_{B}} + {n}_{ii}^{BA_1}+{n}_{ii}^{BA_2}+2{n}_{ii}^{BA_1A_2}\right) \\
      & +\sum_{1\leq j < q \leq N_\mr{p}}\left({n}_{jq}^{B;A_1}+ {n}_{jq}^{B;A_2} + {n}_{jq}^{B_1;B} + {n}_{jq}^{B_2;B}\right) \\
      & + 2 \left({n}_{jq}^{B_1;A_2B} + {n}_{jq}^{B_2;A_1B} + {n}_{jq}^{B;A_1A_2} + {n}_{jq}^{B_1;A_2B}  + {n}_{jq}^{B_1B;A_2} + {n}_{jq}^{B_2A_1;B}\right) .
\end{aligned}
\end{equation}
Importantly the previous expression can be derived using the general results provided by Ref.~\cite{DiMicheleJCP2016} avoiding a direct calculation of ${\cal F}_\mr{multi}$.
 
 \subsection{Mean-field Estimation of the Multivalent Free-Energy}
  
 We now use the multivalent free-energy to calculate the gas-solid phase boundary of particles without substrate (see Main Fig. 2a). We employ a cell model to balance the entropic penalty of caging the colloid into the sites of the solid structure with the multivalent free-energy gain due to inter-particle bridge formation. We consider infinite aggregates with a fixed coordination number, $z$, with $z\leq 6$ as the particles tend to sediment and form bidimensional structures. We estimate the multivalent free-energy gain {\em per} particle, $\Delta F$, by placing all neighboring particles at a fixed distance $d$. In these conditions, all particles feature the same number of bonds, and $\Delta F$ reads as follows (see Eq.~\ref{Eq:FreeEn}) 
\begin{eqnarray}
{ \Delta F \over k_B T } &=& { F_\mr{multi}(d) - F_\mr{multi}(\infty) \over N_p} + F_\mathrm{rep}
\nonumber \\
&=&N_{A_1} \log {n_{A_1}\over n^{(0)}_{A_1}} + N_{A_2} \log {n_{A_2}\over n^{(0)}_{A_2}} + N_{B} \log {n_{B}\over n^{(0)}_{B}}  + F_\mathrm{rep}
 \label{Eq:DeltaF} \\
&& + n_{\mathrm{loop};2} - n^{(0)}_{\mathrm{loop};2} + 2 n_{\mathrm{loop};3} - 2 n^{(0)}_{\mathrm{loop};3} + {1\over 2} \left(  n_{\mathrm{bridge};2} + 2 n_{\mathrm{bridge};3} \right) 
\nonumber 
\end{eqnarray}
where $n_{\mathrm{loop};i}$ and $n_{\mathrm{bridge};i}$ are the total number of bridges and loops formed by $i$ linkers ($i=1,\,2$). The 1/2 factor in front of $ n_{\mathrm{bridge};i}$ accounts for the fact that bridges are shared between two colloids. $n^{(0)}_X$ ($X=A_1,\, A_2,\, B$) and $n^{(0)}_{\mathrm{loop},i}$ are the numbers of free linkers and loops present on isolated particles in the gas phase. In particular, we subtract to $\Delta F$ the contributions of the loops featured by the colloids in the gas phase ($F_\mr{multi}(d=\infty)$, where $d$ is the particle-particle distance). We calculate $n_{\mathrm{loop};i}$ and $n_{\mathrm{bridge};i}$ using Eqs.~\ref{Eq:intra-particle}, \ref{Eq:inter-particle}  ($n^{(0)}_{\mathrm{loop};2}$ and $n^{(0)}_{\mathrm{loop};3}$ follows from the same set of equations with $q_b=0$ and $d=\infty$)
\begin{eqnarray}
n_{\mathrm{loop};2} &=& q_l(d) n^\mathrm{B}(e^{-\beta \Delta G^{BA_1}_0} n^{B_1}+e^{-\beta \Delta G^{BA_2}_0} n^{A_2})
\\
n_{\mathrm{bridge};2} &=& q_b(d) n^{B}z(2 e^{-\beta \Delta G^{BA_1}_0} n^{B_1}+2 e^{-\beta \Delta G^{BA_2}_0} n^{B_2})
\\
n_{\mathrm{loop};3} &=& n^{B_1} n^{B_2}n^{B} q_l(d)^2 e^{-\beta \Delta G^{BA_1A_2}_0}
\\
n_{\mathrm{bridge};3} &=&6 z n^{B_1}n^{B_2}n^{B} q_l(d) q_b(d) e^{-\beta \Delta G^{BA_1A_2}_0} 
\label{Eq:BridgesLoopsMF}
\end{eqnarray}
where we used the fact that $n^{A_1}$, $n^{A_2}$, and $n^{B}$ are the same on all particles (given that each particle interacts with a fixed number of particles, $z$, placed at a fixed distance $d$) and that there are $6\cdot z$ different types of bridges made of three linkers. In particular  
\begin{eqnarray}
n_{\mathrm{bridge};3} &=& z \cdot (n^{A_1;A_2B}_{ij}+n^{A_2;A_1B}_{ij}+n^{B;A_1A_2}_{ij}+n^{A_1A_2;B}_{ij} + n^{B A_1;A_2}_{ij} + n^{BA_2;A_1}_{ij}).
\end{eqnarray}
Notice that from Eq.~\ref{Eq:BridgesLoopsMF} it follows that all types of trimers forming bridges are equally expressed by the system.
\\

In Eq.~\ref{Eq:DeltaF}, $F_\mathrm{rep}$ is a repulsive term accounting for the reduction of the configurational volume available to linkers compressed by pairs of colloids. Neglecting excluded volume interactions between linkers\cite{Leunissen_JCP_2011,Bachmann_SoftMatt_2016,di2018steric} we can write
\begin{eqnarray}
F_\mathrm{rep} = (N_{A_1} +N_{A_2} + N_B) f (L_R,R,d) + N_I f_{\mathrm{rep},I} \, .
\label{Eq:Frep}
\end{eqnarray}
The reactive linkers can be modeled as thin, rigid rods as their length, $L_R$, is much smaller than the persistence length of the dsDNA, $\xi$. The same considerations that led to the calculation of the configurational cost of forming bridges and loop in the previous section can be used to calculate  the entropy reduction of the single reactive linker as follows
\begin{eqnarray}
f (L_R,R,d)= k_B T \log {\Omega_0 - z \cdot v(d,L_R+R,R) \over \Omega_0}
\label{Eq:ExclVolRod}
\end{eqnarray}
where $\Omega_0$ is the space available to the tip of the linkers tethered to isolated colloids ($\Omega_0=4 \pi R^2 L_R$) and $v$ has been defined in Eq.~\ref{Eq:overlapvolume1}.
\\
The inert constructs are longer than the reactive linkers ($L_I\approx 2 L_R$, $L_I\approx \xi/2$)  and are therefore semiflexible. The following equation (with $k=15.1589$, $m=10.3002$, and $\beta=84.85105$) approximates the distribution of the end-to-end distance, ${\bf r}$, of semiflexible filaments with $L=0.5\xi$ (see Fig.~\ref{Fig:TheoryPot})\cite{hamprecht2005end}
\begin{eqnarray}
P_L({\bf r}) \sim \left( r \over L\right)^{k+2} \left[ 1-  \left( r \over L\right)^\beta \right]^m
\, .
\end{eqnarray}
As done for rigid linkers, we approximate the configurational volume reduction with the Euclidean volume excluded to the tip of the semiflexible construct by the presence of the facing particle. This volume reads as the volume excluded to the tip of a rigid rod of length $r$ (Eq.~\ref{Eq:ExclVolRod}) weighted by $\mathrm{P}_L(r)$
\begin{eqnarray}
\Omega^\mathrm{ovl}_I = { \int \mathrm{d} r \cdot \mathrm{P}_L(r) v(d,r+R,R) \over \int_0^L \mathrm{d} r \cdot \mathrm{P}_L(r) }
\label{Eq:VolExclI_1}
\end{eqnarray}
Notice that in the previous equation, the possible orientations of the construct contribute to the calculation of $v$ while $\mathrm{P}_L(r)/\int_0^L \mathrm{d} r \cdot \mathrm{P}_L(r)$ is the probability of having a given end-to-end distance at a given construct direction. We can further simplify Eq.~\ref{Eq:VolExclI_1} by noticing that $v$ is a cubic function in $L$, $R$, and $r$. In the limit in which $r/d,\,r/R \ll 1$ we have that only the liner term in $r$ contributes to $v$. It follows that $\Omega^\mathrm{ovl}_I=v(\langle r \rangle +R,R,d)$, where we defined (see Eq.~\ref{Eq:VolExclI_1}) 
\begin{eqnarray}
\langle r \rangle = { \int_0^L \mathrm{d} r \cdot \mathrm{P}_L(r) \cdot r  \over \int_0^L \mathrm{d} r \cdot \mathrm{P}_L(r) } =0.922\cdot L_I\, . 
\end{eqnarray}
Finally the repulsive contribution {\em per} inert construct (see Eq.~\ref{Eq:Frep}) reads as follows 
\begin{eqnarray}
 f_{\mathrm{rep},I} =  f (0.922\cdot L_I,R,d)
\end{eqnarray}

\subsection{Calculation of the phase boundary} 

For square-well potentials with well depth and width equal, respectively, to $\epsilon$ and $\sigma$, the phase boundary satisfies the following equation \cite{sear1999stability,charbonneau2007gas}
\begin{eqnarray}
\beta \epsilon = \log \left({\rho \delta^3\over 8} \right)  
\label{Eq:CellModel}
\end{eqnarray}
where $\rho$ is the density of the particles in the fluid phase. To use Eq.~\ref{Eq:CellModel}, we map the free energy profiles as a function of the interparticle distance, $\Delta F(d)$, into square well potentials as follows (see Fig.~\ref{Fig:TheoryPot}):
\begin{itemize} 
\item{We identify the width of the well with the minimum of the multivalent free energy $\epsilon=\Delta F(d_\mathrm{min})$.}  
\itemize{The two boundaries ($x_\pm$) of the square well are identified with the distances at which the multivalent free-energy is half the value of $\Delta F(d_\mathrm{min}$), $\Delta F(x_\pm)=\Delta F(x_\mathrm{min})/2$. It follows that $\delta=x_\mathrm{min}-x_\mathrm{max}$.}
\end{itemize}

Notice that the profile of $\Delta F(d)$ is a function of the particle density ($\rho$), the temperature ($T$), the valency of the aggregate ($z$), and the fraction of linkers $f$ (see main text). In particular, inert constructs sensibly increase the value of $d_\mathrm{min}$, reducing the width of the well, $\delta$. Therefore when changing, for instance, the number of reactive linkers to find the value of $f$ at coexistence, one should also change the values of $d_\mathrm{min}$ (used to calculate $\epsilon$) and $\delta$ in Eq.~\ref{Eq:CellModel}. Practically, we start with an initial guess for $d_\mathrm{min}$ and $\delta$, calculate the phase boundary using Eq.~\ref{Eq:CellModel}, adjust the well parameters ($d_\mathrm{min}$ and $\delta$) using $\Delta F(d)$ at the coexistence point, and recalculate the phase boundary and the well parameters until reaching convergence.\\
The phase boundary is calculated for $z=4, 5, 6$, and a particle packing fraction $\phi=0.28\%$, $2.8\%$, $28\%$, both ranges comfortably encompassing the coordination observed in experimental aggregates and the experimental packing fraction. As discussed in Sec~\ref{chamebrs}, $\phi\sim 3-5\%$ as estimated near the bottom of the experimental cell accounting for particle sedimentation. The values of $d_\mathrm{min}$ and $\delta$ corresponding to the tested conditions are summarised in Tab.~\ref{tabletheory}. Because the well parameters are weakly affected by the temperature (see Fig.~\ref{Fig:TheoryPot}), we use the same square well to model $\Delta F(d)$ at different temperatures. Figure~\ref{FigureS8} shows a zoomed-in view of the computed phase boundaries, demonstrating the relatively weak dependence on $z$ and $\phi$. The expanded phase boundary shown in Fig.~2 conservatively accounts for the entire range in Fig.~\ref{FigureS8}.

\clearpage

\begin{center}
\begin{table}[h!]
\begin{tabular}{ p{1.5cm} p{14cm} } 
\hline
   \footnotesize $SEA_1$ & \footnotesize  \textbf{CCGTTCGC} \emph{TTTT} GGTTTGTTGTTGTGTTGG \\
   \footnotesize $SEA_2$ & \footnotesize  \textbf{TCGCCTGG} \emph{TTTT} GGTTTGTTGTTGTGTTGG \\
   \footnotesize $SEB$ & \footnotesize GTGTTGAGTAGTGAGATG \emph{TTTT} \textbf{CCAGGCGAACGGCGTC}\\
   \footnotesize $SEC$ & \footnotesize GTGTTGAGTAGTGAGATG \emph{TTTT} \textbf{GACGCCGTTCGCCTGG}\\
    \footnotesize $CHA_1$ & \footnotesize GTGTTTGTGGTGTGATTG  (TEG) Cholesterol \\
    \footnotesize $CHA_2$ & \footnotesize Cholesteryl (TEG) CAATCACACCACAAACACCCAACACAACAACAAACC\\
    \footnotesize $CHB_1$ & \footnotesize  CAACATCTCACTACTCAACACCACACTCACCACCACAAC (TEG) Cholesterol\\
    \footnotesize $CHB_2$ & \footnotesize Cholesteryl (TEG)  GTTGTGGTGGTGAGTGTG \\
    \footnotesize $I_1$ & \footnotesize GTGTTGAGTAGTGAGATGCCAACACCACAGATATCACAACCACAACCAAC\\
     \footnotesize $I_2$ & \footnotesize GTTGGTTGTGGTTGTGATATCTGTGGTGTTGG \\
     \footnotesize $Fluo_1$ & \footnotesize Cy5 GGTTTGTTGTTGTGTTGG \\
     \footnotesize $Fluo_2$ & \footnotesize GTGTTGAGTAGTGAGATG Cy3 \\
 \hline

\end{tabular}
 \caption{\textbf{Oligonucleotide sequences.} (TEG): Triethylene glycol. Bases in italic are unpaired, while sticky ends are shown in bold. Domains are separated by spaces. Oligonucleotides $CHA_2$ and $CHB_2$ are purchased from Eurogentec, all other strands from Integrated DNA technologies. Linkers and other constructs are assembled from the following oligonucleotides: $A_1 = SEA_1+CHA_1+CHA_2$; $A_2=SEA_2+CHA_1+CHA_2$; $B=SEB+CHB_1+CHB_2$; $C=SEC+CHB_1+CHB_2$; $I=I_1+I_2+CHB_1+CHB_2$; Cy5-labelled construct $=Fluo_1+CHA_1+CHA_2$; Cy3-labelled construct $=Fluo_2+CHB_1+CHB_2$. { The sequences of the sticky ends were adapted manually from those used in Parolini \emph{et al.}~\cite{parolini2016controlling} Cholesterolised strands $CHA_1$, $CHA_2$, $CHB_1$ and $CHB_2$ were previously used in Kaufhold \emph{et al.}~\cite{Kaufhold:2019aa} The remaining strands and domains were designed and tested with the NUPACK web server.~\cite{zadeh2011nupack}}}
 \label{table}
\end{table}
\end{center}

\clearpage

\begin{center}
\begin{table}[h!]
\begin{tabular}{|l|l|l|l|} 
\hline
   packing fraction ($\phi$) & valency ($z$) &  $d_\mathrm{min}$  & $\delta$  \\
\hline
  0.28 & 4 & 1019.5$\,$nm &  2.9$\,$nm \\ 
  0.28 & 5 &1019.7$\,$nm &  2.6$\,$nm \\
  0.28 & 6 & 1019.8$\,$nm &  2.4$\,$nm \\  
 \hline
  0.028 & 4 & 1019.4$\,$nm &  3.05$\,$nm \\ 
  0.028 & 5 & 1019.62$\,$nm &  2.725$\,$nm \\
  0.028 & 6 & 1019.75$\,$nm &  2.525$\,$nm \\  
 \hline
  0.0028 & 4 & 1019.33$\,$nm &  3.25$\,$nm \\ 
  0.0028 & 5 & 1019.55$\,$nm &   2.875 \\
  0.0028 & 6 & 1019.69$\,$nm &  2.625$\,$nm \\  
 \hline
\end{tabular}
 \caption{Square-well parameters used in Eq.~\ref{Eq:CellModel} to calculate the phase boundary.}
 \label{tabletheory}
\end{table}
\end{center}

\clearpage

\clearpage

\begin{figure}[ht!]
\centering
 \includegraphics[width=17cm]{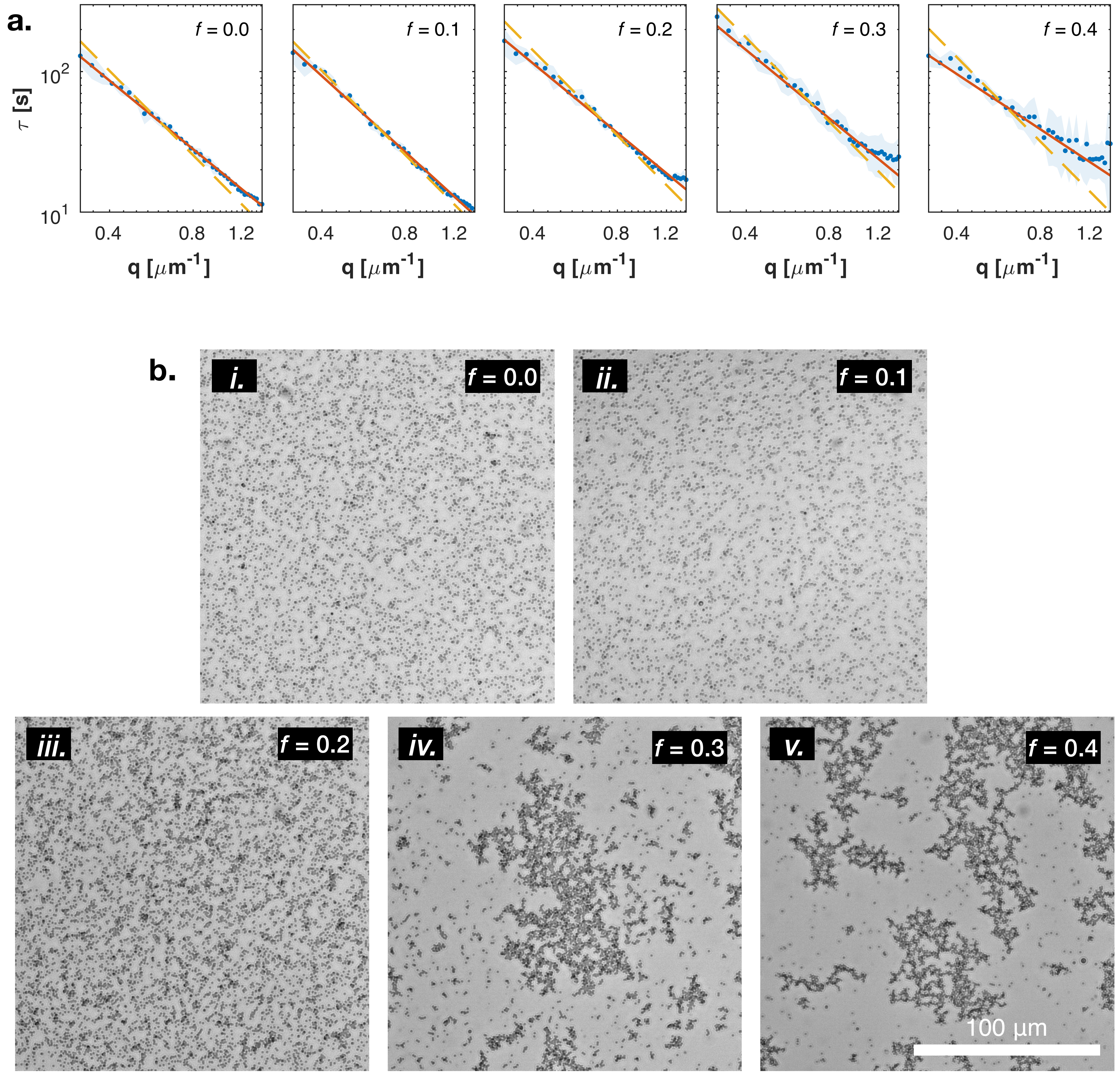}
  \caption{\textbf{Assessing particle aggregation visually and \emph{via} DDM}. \textbf{a.} Experimental values of the DDM relaxation time $\tau$ as a function of the wave vector $q$ recorded at the end of an aggregation experiment ($t=22$ hours) for all tested values of the fraction of linkers $f$. Points and the surrounding shaded region indicate, respectively, the mean and standard deviation calculated over 8 ROIs (2 fields of view). The solid line indicates the best power law fit $\tau \propto q^{-\alpha}$, while the dashed line the best Brownian fit $\tau = D q^{-2}$. The latter is used to extract the effective diffusion coefficient $D$, shown in Fig. 2\textbf{b} and Fig.~\ref{FigureS3}. Note that the datapoints deviate more significantly from the Brownian slope at large $f$, following the formation of branched aggregates with a complex dynamics.~\cite{Cho:2020aa}. \textbf{b.} Bright field microscopy snapshots from the movies underlying the DDM data in panel \textbf{a}.}
  \label{FigureS2}
\end{figure}

 \clearpage
 
 \begin{figure}[ht!]
\centering
  \includegraphics[width=9cm]{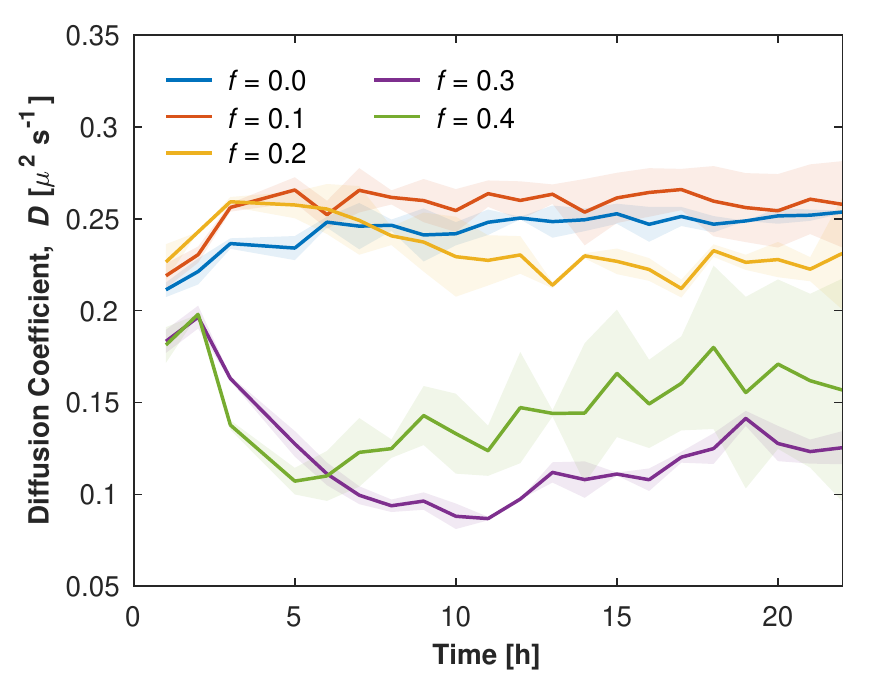}
  \caption{\textbf{Time evolution of the DDM effective diffusion coefficient for samples featuring both particles and substrate spheres.} Note the similarity with the curves in Fig.~2\textbf{b}, indicating that the bulk phase behaviour of particles is unaffected by the substrate spheres, which have the only effect of regulating the deposition of some particles on their surface. { The slight increase in $D$ observed at the beginning of the experiment in all sample may be a consequence of initial thermalisation.}}
  \label{FigureS3}
\end{figure}

\clearpage

\begin{figure}[ht]
\begin{center}
\includegraphics[width=10cm]{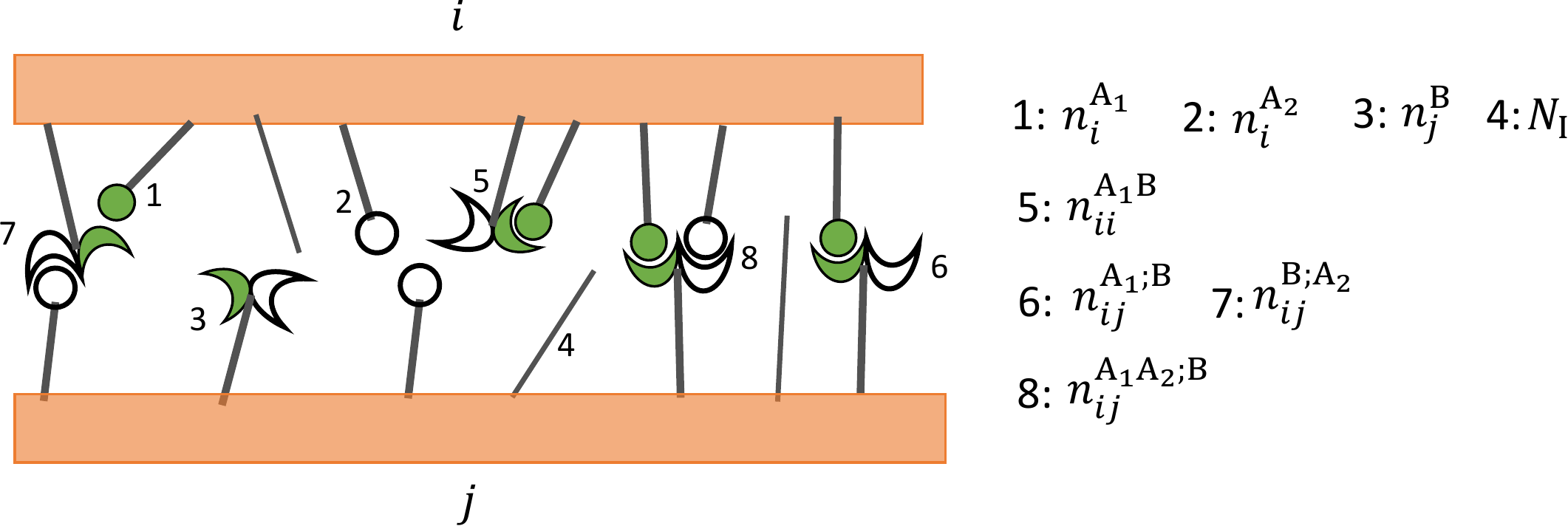}
\caption{\textbf{Examples of intra-particle and inter-particle complexes.} The planes represent the surface of particles $i$ and $j$ and carry reactive (${A_1}$, ${A_2}$, and ${B}$)  and inert (I) linkers. $n^\mr{X}_p$ denotes the number of free linkers of type X ($\mr{X}={A_1}$, ${A_2}$, or ${B}$) tethered to particle $p$. Each complex is identified by its monomeric components and the planes to which they are anchored. For bridges, semicolumns separate the components tethered to particle $i$ from those tethered to particle $j$. Each particle carries $N_I$ inert constructs. }
\label{Fig:complexes}
\end{center}
\end{figure}

\clearpage

\begin{figure}[t]
\begin{center}
\includegraphics[width=8cm]{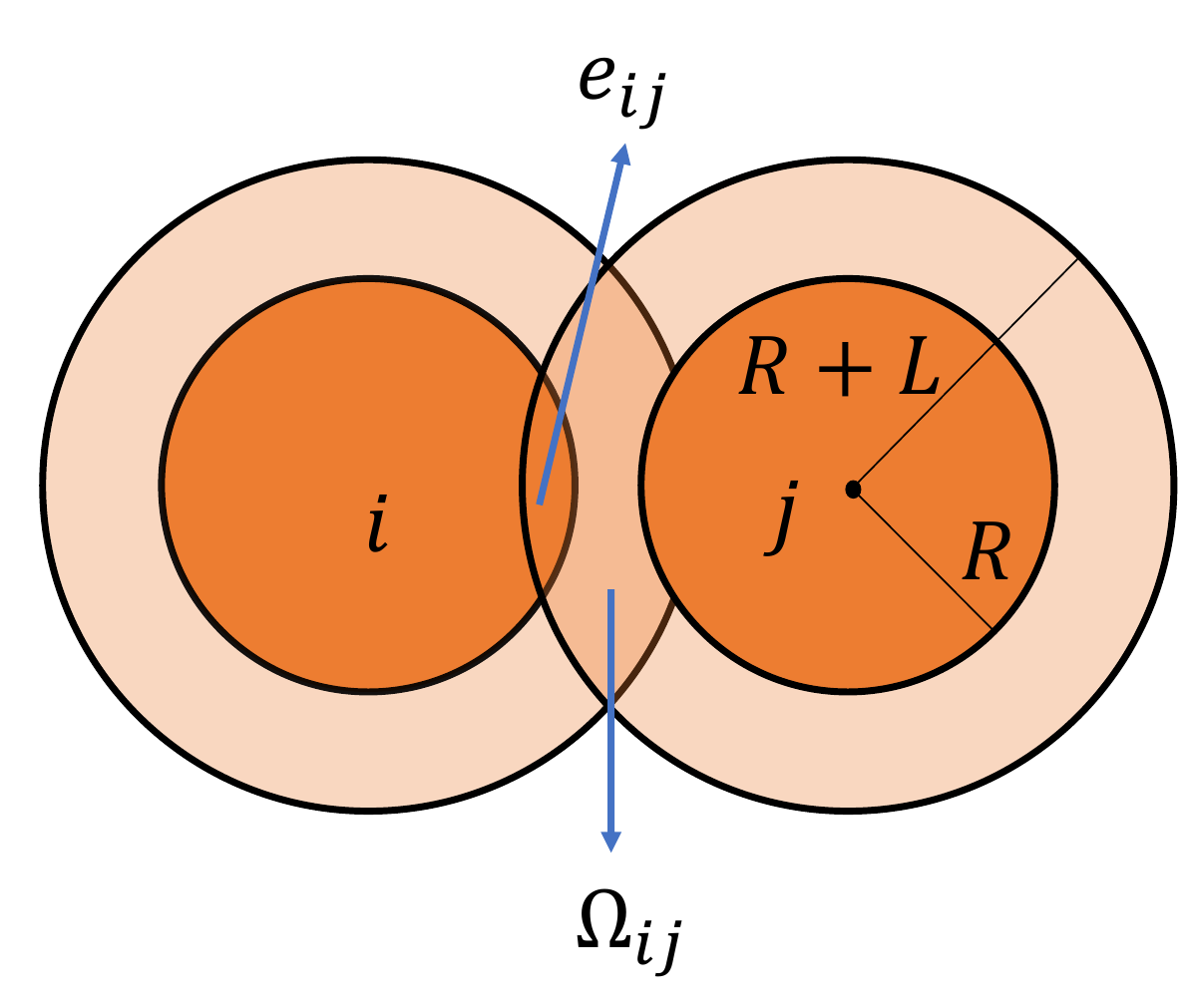}
\caption{\textbf{Configurational volumes.} Configurational volume excluded to a linker tethered to particle $j$ by the presence of particle $i$ ($e_{ij}$) and configurational volume available to interparticle bridges ($\Omega_{ij}$). The definitions of $e_{ij}$ and $\Omega_{ij}$ are given in Eq. \ref{Eq:overlapvolume}. $R$ and $L$ denote the radius of the particles and the length of the linkers, respectively.}
\label{Fig:overlapvolume}
\end{center}
\end{figure}

 \clearpage

\begin{figure}[ht!]
\centering
  \includegraphics[width=9cm]{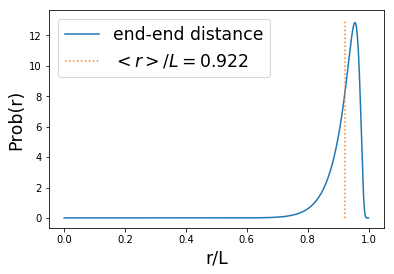}
  \caption{\textbf{Distribution of the end-to-end distance of a semiflexible rod} with persistence length equal to twice the length of the rod $L$ (from \cite{hamprecht2005end}). The dotted line nicks the average distance with a fixed end-to-end direction.  }
  \label{Fig:TheoryPot}
\end{figure}

\clearpage

\begin{figure}[ht!]
\centering
  \includegraphics[width=9cm]{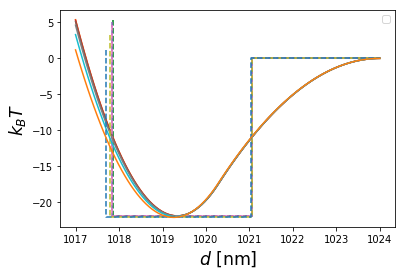}
  \caption{{\bf Mapping free-energy profiles into square-well potentials.} Full lines represent the multivalent free energies $\Delta F$ calculated using Eq.~\ref{Eq:DeltaF} while dashed lines the corresponding square-well potentials (see text). Different colors represent different temperatures ($T=20\,^\circ$C, $26\,^\circ$C, $T=33\,^\circ$C, $39\,^\circ$C, $45\,^\circ$C, and $50\,^\circ$C). Valency is equal to $z=4$ and the packing fraction to $\phi=28\%$. }
  \label{Fig:TheoryPot}
\end{figure}

\clearpage

\begin{figure}[ht!]
\centering
  \includegraphics[width=12cm]{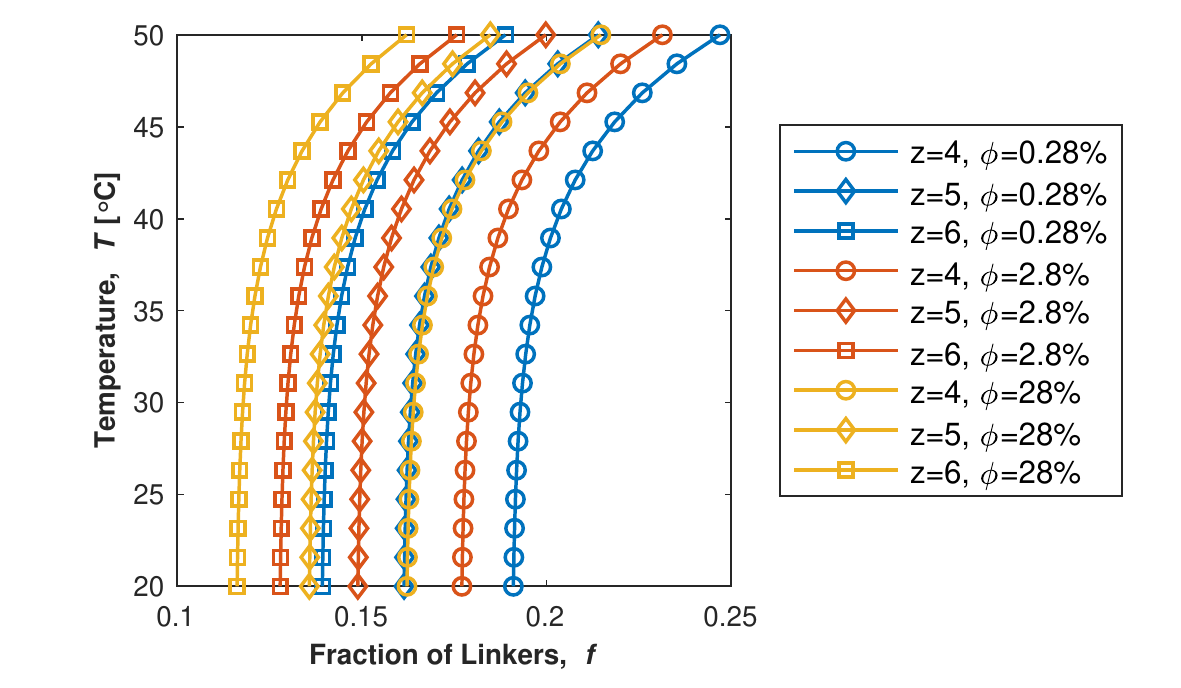}
  \caption{{\bf Liquid-solid phase boundaries} as calculated using the parameters in Table~\ref{tabletheory}\label{FigureS8}.}
  \label{Fig:TheoryPot}
\end{figure}

\clearpage